\documentclass{emulateapj}
\usepackage{apjfonts}

\newcommand{\etal}{et~al.}
\newcommand{\MgIIdblt}{{\rm Mg}\kern 0.1em{\sc ii}~$\lambda\lambda 2796, 2803$}
\newcommand{\MgII}{\hbox{{\rm Mg}\kern 0.1em{\sc ii}}}
\newcommand{\HI}{\hbox{{\rm H}\kern 0.1em{\sc i}}}
\newcommand{\CIV}{\hbox{{\rm C}\kern 0.1em{\sc iv}}}
\newcommand{\OVI}{\hbox{{\rm O}\kern 0.1em{\sc vi}}}
\newcommand{\Lya}{\hbox{{\rm Ly}\kern 0.1em $\alpha$}}
\newcommand{\kms}{\hbox{km~s$^{-1}$}}

\newcommand{\frl}{\hbox{$f_{\hbox{\tiny $R(L)$}}$}}
\newcommand{\fdmax}{\hbox{$f_{\hbox{\tiny $D$max}}$}}
\newcommand{\fdbar}{\hbox{$f_{\hbox{\tiny $\langle D\, \rangle$}}$}}
\newcommand{\magiicat}{\hbox{{\rm MAG}{\sc ii}CAT}}

\shorttitle{\sc The {\MgII} Circumgalactic Medium}
\shortauthors{\sc Nielsen, Churchill, \& Kacprzak}
\slugcomment{Accepted for publication in ApJ, August 5, 2013}

\begin{document}

\title{~{\magiicat} II. General Characteristics of the \\
~{\MgII} Absorbing Circumgalactic Medium}

\author{\sc
Nikole M. Nielsen\altaffilmark{1},
Christopher W. Churchill\altaffilmark{1},
and
Glenn G. Kacprzak\altaffilmark{2,3}
}
\altaffiltext{1}{New Mexico State University, Las Cruces, NM 88003
{\tt nnielsen@nmsu.edu, cwc@nmsu.edu}}
\altaffiltext{2}{Swinburne University of Technology, Victoria 3122,
Australia {\tt gkacprzak@astro.swin.edu.au}}
\altaffiltext{3}{Australian Research Council Super Science Fellow}

\begin{abstract}

We examine the {\MgII} absorbing circumgalactic medium (CGM) for the
182 intermediate redshift ($0.072\leq z \leq 1.120$) galaxies in the
``{\MgII} Absorber-Galaxy Catalog'' ({\magiicat}, Nielsen {\etal}). We
parameterize the anti-correlation between equivalent width,
$W_r(2796)$, and impact parameter, $D$, with a log-linear fit, and
show that a power law poorly describes the data. We find that higher
luminosity galaxies have larger $W_r(2796)$ at larger $D$
($4.3~\sigma$). The covering fractions, $f_c$, decrease with
increasing $D$ and $W_r(2796)$ detection threshold. Higher luminosity
galaxies have larger $f_c$; no absorption is detected in lower
luminosity galaxies beyond $100$~kpc. Bluer and redder galaxies have
similar $f_c$ for $D<100$~kpc, but for $D>100$~kpc, bluer galaxies
have larger $f_c$, as do higher redshift galaxies. The ``absorption
radius,'' $R(L)=R_{\ast}(L/L^{\ast})^{\beta}$, which we examine for
four different $W_r(2796)$ detection thresholds, is more luminosity
sensitive to the $B$-band than the $K$-band, more sensitive for redder
galaxies than for bluer galaxies, and does not evolve with redshift
for the $K$-band, but becomes more luminosity sensitive towards lower
redshift for the $B$-band. These trends clearly indicate a more
extended {\MgII} absorbing CGM around higher luminosity, bluer, and
higher redshift galaxies. Several of our findings are in conflict with
other works. We address these conflicts and discuss the implications
of our results for the low-ionization, intermediate redshift CGM.

\end{abstract}


\section{Introduction}
\label{sec:intro}

Understanding the formation and evolution of galaxies is one of the
foremost problems facing extragalactic research. Over the last decade,
theoretical activity has been focused on studying how galaxies form
and evolve in the context of the response of baryonic gas to dark
matter halos of various masses and to physical processes such as
stellar and AGN feedback \citep[e.g.,][]{birnboim03, maller04,
  keres05, dekel06, birnboim07, ocvirk08, keres09, oppenheimer10,
  stewart11, vandevoort11, vandevoort+schaye11}. These theoretical
works have established that the circumgalactic medium (CGM) is a
dynamic environment comprising inflowing accretion, outflowing winds,
and gas that recycles between the halo and galaxy.

What has become clear is that the CGM is intimately linked to galaxy
morphology, stellar populations and kinematics, and chemical
evolution. The detailed physics governing the heating and cooling of
baryonic gas regulates the formation and galactic-scale dynamical
motions of stars, which in turn feedback and govern the physics of the
gas in the CGM \citep[e.g.,][]{dekel86, ceverino09}. Since this
paradigm of galaxy evolution suggests a strong CGM-galaxy connection,
many investigators have taken a phenomenological approach to
developing models that predict the observable geometric distribution
and kinematics of baryonic gas in and around galaxies
\citep[e.g.,][]{weisheit78, lb92, cc96, mo96, benjamin97, cc98,
  tinkerchen08, chelouche10, bouche11}.

As such, the CGM is an astrophysical environment which, at any point
in the evolution of a galaxy, yields clues to both its historical
development and future evolution. Therefore, observations of the CGM
around individual galaxies promise to provide highly sought
constraints on the physics governing the global properties of
galaxies. In practice, the CGM is almost exclusively accessible via
absorption lines present in the spectra of background objects
(traditionally using the technique of quasar absorption lines).
Studying the CGM-galaxy connection is desirable over all redshifts,
with the vast majority of work to date having been focused on $z \leq
1$ \citep[however, cf.,][]{adelberger05, simcoe06, steidel10,
  rudie12}. At these redshifts, galaxy photometric and spectral
properties can be measured with relatively high accuracy and modest
investment in telescope time with {\MgII} absorption
\citep[e.g.,][]{bb91, sdp94, cwc-china, chen10a, kacprzak}.

The {\MgIIdblt} doublet is especially well suited at intermediate
redshifts since the absorption is observable in the
optical. Furthermore, {\MgII} absorption is ideal for tracing low
ionization, metal-enriched CGM gas in that it arises in a wide range
of {\HI} environments, from $\log N({\HI}) \simeq 16.5$ to greater
than 21.5 \citep{bs86, ss92, weakI, archiveI, rao00, weakII}. It is
also well established that {\MgII} absorption probes key CGM
structures such as outflow gas \citep[e.g.,][]{tremonti07, weiner09,
  martin09, rubin10} and inflow gas \citep[e.g.,][]{ribaudo11,
  rubin11, thom11, ggk-1317} associated with galaxies.

Through the efforts of the aforementioned studies, the behavior of
{\MgII} CGM absorption in relation to various galaxy properties has
been incrementally characterized over the last two decades. Examples
of some of the findings include relationships between the strength of
{\MgII} absorption and galaxy luminosity \citep{ggk08, chen10a},
galaxy mass \citep{bouche06, gauthier09, churchill-masses}, star
formation or specific star formation rate \citep{chen10b, menard11},
galaxy color \citep{zibetti07, bordoloi11}, galaxy morphology
\citep{ggk-morphology}, and galaxy orientation \citep{bordoloi11,
  bouche11, kacprzak, kcn, churchill-weakgals}.

Generally speaking, these investigations have yielded results that
appear to be converging on an observational portrait of the {\MgII}
absorbing CGM that is more or less consistent with the broad scenario
forwarded by theory and simulations. For example, evidence is mounting
that accretion may preferentially reside in a coplanar geometry
\citep[e.g.,][]{steidel97, ggk-sims, stewart11, kacprzak}, whereas
winds may outflow along the galaxy minor axis \citep{bordoloi11,
  bouche11, churchill-weakgals, kcn}. However, due to the complex
ionization structure, non-uniform metal enrichment, and dynamical
processes of the CGM, and due to the wide range of galaxy properties,
such as luminosity, color, mass, and morphology, the data exhibit
substantial scatter. As such, while some works have found
statistically significant correlations between two quantities or
between combined/scaled quantities, many works have reported
connections between galaxy and {\MgII} absorption CGM properties
either based on general trends (i.e., correlations that are not
statistically significant above the $3~\sigma$ level) or on slight
reductions in the scatter of certain relationships by
combining/scaling one or more measured quantity.

We compiled the ``{\MgII} Absorber-Galaxy Catalog'' ({\magiicat})
which is described in detail in Paper I of this series
\citep{nielsen12a}, with the aim to further illuminate the CGM-galaxy
connection at higher statistical significance and over a wide range of
galaxy properties. In this paper, we utilize the data to address
several long-standing questions with regard to the {\MgII} CGM-galaxy
connection. In \S~\ref{sec:sample}, we briefly describe the galaxy
sample, which has uniform photometric properties and impact parameters
for a $\Lambda$CDM cosmology ($H_0=70$ km~s$^{-1}$~Mpc$^{-1}$,
$\Omega_M=0.3$ and $\Omega_{\Lambda}=0.7$). In \S~\ref{sec:results},
we present an examination of (1) galaxy color and absolute magnitude
evolution, (2) the dependence of $W_r(2796)$ on galaxy color, (3) the
dependence of $W_r(2796)$ on impact parameter, (4) covering fraction
as a function of impact parameter and $W_r(2796)$ for luminosity,
color, and redshift subsamples, and (5) the luminosity scaling of
``halo absorption radius'' and the behavior of covering fraction with
$W_r(2796)$, galaxy color, and redshift. In \S~\ref{sec:discussion},
we discuss the implications of our results and give concluding remarks
in \S~\ref{sec:concl}.

\section{Galaxy Sample and Subsamples}
\label{sec:sample}

From an exhaustive literature search, we compiled a catalog of
galaxies with spectroscopic redshifts $0.07 \leq z \leq 1.1$ within a
projected distance of $D \leq 200$~kpc from a background quasar, with
known {\MgII} absorption or an upper limit on absorption less than
0.3~{\AA}. We distinguish between isolated and group galaxies, where
group galaxies have a nearest neighbor within 100~kpc and have a
velocity separation no greater than 500 {\kms}, and focus only on
isolated galaxies in the present work. The total catalog consists of
182 isolated galaxies. Full details of all galaxies and {\MgII}
absorption properties as well as the various selection methods used
are presented in Paper I. Here we briefly describe the general
properties of {\magiicat}.

\begin{deluxetable*}{ccccccccccccccc}
\tablecolumns{15}
\tablewidth{0pt}
\tablecaption{Subsample Characteristics \label{tab:sschars}}
\tablehead{
\colhead{\phantom{x}} &
\colhead{\phantom{x}} &
\colhead{\phantom{x}} &
\multicolumn{3}{c}{----------- $z_{\rm gal}$ ------------} &
\multicolumn{3}{c}{---------- $L_B/L_B^{\ast}$ ----------} &
\multicolumn{3}{c}{---------- $L_K/L_K^{\ast}$ ----------} &
\multicolumn{3}{c}{---------- $B-K$ ----------} \\
\colhead{Subsample} & 
\colhead{\# Gals}   & 
\colhead{Cut\tablenotemark{a}} &  
\colhead{Min} &
\colhead{Max} &
\colhead{Mean}     &  
\colhead{Min} &
\colhead{Max} &
\colhead{Mean} & 
\colhead{Min\tablenotemark{b}} &
\colhead{Max\tablenotemark{b}} &
\colhead{Mean\tablenotemark{b}} &
\colhead{Min\tablenotemark{b}} &
\colhead{Max\tablenotemark{b}} &
\colhead{Mean\tablenotemark{b}}        
}
\startdata
All                 & 182 & $\cdots$ & 0.072 & 1.120 & 0.418 & 0.017 & 5.869 & 0.855 & 0.006 & 9.712 & 0.883 & 0.038 & 4.090 & 1.537 \\[3pt]
Low z               &  91 & 0.359 & 0.072 & 0.358 & 0.225 & 0.017 & 3.759 & 0.734 & 0.006 & 4.901 & 0.654 & 0.283 & 3.037 & 1.554 \\[3pt]
High z              &  91 & 0.359 & 0.359 & 1.120 & 0.612 & 0.071 & 5.869 & 0.976 & 0.016 & 9.712 & 1.123 & 0.038 & 4.090 & 1.520 \\[3pt]
Low $L_B/L_B^{\ast}$  &  91 & 0.611 & 0.072 & 0.941 & 0.411 & 0.017 & 0.610 & 0.306 & 0.006 & 1.680 & 0.271 & 0.038 & 4.090 & 1.410 \\[3pt]
High $L_B/L_B^{\ast}$ &  91 & 0.611 & 0.096 & 1.120 & 0.425 & 0.613 & 5.869 & 1.405 & 0.241 & 9.712 & 1.452 & 0.483 & 3.303 & 1.655 \\[3pt]
Low $L_K/L_K^{\ast}$  &  82 & 0.493 & 0.110 & 0.941 & 0.396 & 0.019 & 1.440 & 0.376 & 0.006 & 0.487 & 0.222 & 0.038 & 3.037 & 1.231 \\[3pt]
High $L_K/L_K^{\ast}$ &  82 & 0.493 & 0.096 & 1.017 & 0.410 & 0.189 & 5.869 & 1.359 & 0.499 & 9.712 & 1.544 & 0.637 & 4.090 & 1.843 \\[3pt]
Blue                &  82 & 1.482 & 0.098 & 1.017 & 0.444 & 0.019 & 3.522 & 0.803 & 0.006 & 2.602 & 0.454 & 0.038 & 1.478 & 1.047 \\[3pt]
Red                 &  82 & 1.482 & 0.096 & 0.852 & 0.362 & 0.035 & 5.869 & 0.931 & 0.026 & 9.712 & 1.312 & 1.487 & 4.090 & 2.027 \\[-5pt]
\enddata 
\tablenotetext{a}{The median value by which each subsample was
  bifurcated. The cut is inclusive to the ``high'' bins and red bin.}
\tablenotetext{b}{Only including galaxies with a value for $M_K$.}
\end{deluxetable*}

The sample presented here comprises 182 isolated galaxies toward 134
sightlines, covering a redshift range $0.072 \leq z_{\rm gal} \leq
1.120$, with a median of $\langle z \rangle =0.359$. The impact
parameter range is $5.4 \leq D \leq 194$~kpc, with median $\langle D
\rangle =49$~kpc.

For each galaxy, we determined rest-frame AB absolute magnitudes,
$M_B$ and $M_K$, by computing $k$-corrections appropriate for the
observed magnitudes \citep[e.g.,][]{kim} using the extended
\citet{cww80} spectral energy distribution (SED) templates from
\citet{bolzonella00}. We selected a galaxy SED by comparing the
observed galaxy colors to the redshifted SEDs. Luminosities,
$L_B/L_B^{\ast}$ and $L_K/L_K^{\ast}$, were obtained using a linear
fit to $M_B^{\ast}$ with redshift from \citet{faber} ($B$-band) and
using $M_K^{\ast}(z)$ as expressed in Eq.~2 from \citet{cirasuolo}
($K$-band). Absolute $B$-band magnitudes range from $-16.1 \geq M_B
\geq -23.1$, corresponding to luminosities of $0.02 \leq
L_B/L_B^{\ast} \leq 5.87$, with median
$L_B/L_B^{\ast}=0.611$. Absolute $K$-band magnitudes range from $-17.0
\geq M_K \geq -25.3$, corresponding to $0.006 \leq L_K/L_K^{\ast} \leq
9.7$. Rest-frame $B-K$ colors range from $0.04 \leq B-K \leq 4.09$,
with median $B-K=1.48$. We obtain $K$-band absolute magnitudes and
luminosities, and $B-K$ colors for all but 18 of the galaxies.

Galaxies in {\magiicat} were bifurcated into several subsamples for
analysis. When dividing the galaxy-absorber pairs into subsamples, we
either took a sample driven approach by splitting the sample at the
median value of a given observed quantity, or dividing the sample
based upon historical precedent, such as $W_r(2796)$ cuts. The full
sample was sliced by the median galaxy redshift, $L_B/L_B^{\ast}$,
$L_K/L_K^{\ast}$, or $B-K$. Table~\ref{tab:sschars} presents the
characteristics of the full sample and each subsample, including the
number of galaxies, the median value by which the catalog was cut, and
the minimum, maximum, and mean values of galaxy redshift, $B$- and
$K$-band luminosity, and $B-K$ color. The subsample names listed will
be used throughout this paper.

\section{Results}
\label{sec:results}

In this section, we report on the luminous properties of the galaxies
and directly compare them to the properties of the {\MgII} absorbing
CGM. The analysis presented here is based upon direct observables. We
do not scale or combine measured quantities. We also do not scale or
fit the data to models of the CGM. Our aim is to characterize the
{\MgII} CGM-galaxy connection directly with no assumptions and to
provide information that can be directly interpreted in terms of
primary observables. We examine only bivariate relationships,
reserving further analysis, such as multivariate techniques for future
work to appear in later papers of this series.

\subsection{Galaxy Magnitudes, Luminosities, and Colors\\ Versus Redshift}
\label{sec:zcolors}

Since the CGM as probed by {\MgII} may depend upon galaxy stellar
populations, and these populations are known to evolve, we examined
whether the galaxy magnitudes and rest-frame colors evolve with
redshift for the sample. A Kendall-$\tau$ rank correlation test
yielded a $4.4~\sigma$ significance that $M_B$ correlates with $z_{\rm
  gal}$ (i.e., galaxies at higher redshift are brighter in the
$B$-band), whereas $M_K$ shows a weak trend for redshift evolution
($2.2~\sigma$). \citet{faber} found that $M_B^{\ast}$ brightens with
increasing redshift from the DEEP2+COMBO-17 surveys. The average $M_B$
of the sample presented here increases with redshift more rapidly than
$M_B^{\ast}$. Similarly, \citet{cirasuolo} found that $M_K^{\ast}$
tends to brighten with increasing redshift from UKIDSS, whereas the
average $M_K$ of the sample presented here does not exhibit redshift
evolution. As such, we may be seeing that galaxies associated with
{\MgII} absorption or an upper limit on absorption are characterized
by bluer colors at higher redshift. However, selection effects may be
important. Virtually all the galaxies were selected or discovered
\citep[for a summary of the selection methods used, see][]{nielsen12a}
using the red band in the observer frame, which probes further toward
the $B$-band in the galaxy rest frame with increasing
redshift. However, this cannot explain the lack of evolution in the
$K$-band.

We also conducted a rank correlation test on galaxy luminosities
versus redshift, $z_{\rm gal}$, since we calculate the luminosities
from the absolute magnitudes. In the $B$-band, we find no redshift
evolution of $L_B/L_B^{\ast}$ to the $1.2~\sigma$ level. Similarly in
the $K$-band, we find that $L_K/L_K^{\ast}$ does not correlate with
redshift at the $0.9~\sigma$ level. The lack of correlations here are
due to the fact that we take into account the redshift evolution of
$M_B^{\ast}$ from \citet{faber} and $M_K^{\ast}$ from
\citet{cirasuolo} in our luminosity calculations.

A rank correlation test on $B-K$ versus $z_{\rm gal}$ shows that the
null hypothesis of no correlation can be ruled out to a confidence
level (CL) no better than $1.8~\sigma$, indicating that the $B-K$
rest-frame color of galaxies in {\magiicat} does not evolve with
redshift. This result is consistent with \citet{zibetti07} who use
statistical methods in which individual galaxies were not directly
identified with absorbers for $0.4 \le z \le 1$ using observed $g$,
$r$, $i$, and $z$ band magnitudes for $W_r(2796) \ge 0.8$~{\AA}. A
direct comparison is difficult because we use optical $B$-band and
$2.2\mu$m infrared ($K$-band) absolute magnitudes to determine
rest-frame colors and we study systems with much smaller $W_r(2796)$
than represented by their sample. Slicing the sample presented here at
$W_r(2796)=0.8$~{\AA} to compare to the \citet{zibetti07} sample, we
obtain $0.8~\sigma$ for $W_r(2796) \geq 0.8$~{\AA} and for $W_r(2796)
< 0.8$~{\AA} we obtain $1.7~\sigma$.

\subsection{Galaxy Luminosities and Colors Versus $W_r(2796)$}
\label{sec:wcolors}

To determine whether $W_r(2796)$ exhibits some dependency on galaxy
luminosity, we performed a non-parametric Kendall's $\tau$ rank
correlation test that allows for upper limits on either the dependent
or independent variable for bivariate data \citep[see][]{bhk,wang}
with $W_r(2796)$ as the dependent variable. The null hypothesis of no
correlation could not be ruled out for $W_r(2796)$ against
$L_B/L_B^{\ast}$ ($0.2~\sigma$), $L_K/L_K^{\ast}$ ($0.6~\sigma$),
$M_B$ ($0.9~\sigma$), and $M_K$ ($0.5~\sigma$). The lack of
correlations found here are interesting in view of the arguments by
\citet{bouche06}, who derive an anti-correlation between $W_r(2796)$
and galaxy luminosity with a dependence on the faint-end slope of the
galaxy luminosity function and a proportionality between galaxy
luminosity and the cross-section of the absorbing CGM.

To determine if $W_r(2796)$ has any dependency on galaxy color, we
performed the non-parametric Kendall's $\tau$ rank correlation test on
$W_r(2796)$ against $B-K$. The test indicates that $W_r(2796)$ does
not directly correlate with galaxy color for the full sample
($1.3~\sigma$). A Kendall's $\tau$ rank correlation test between
$W_r(2796)$ and $B-K$ for the subsample with $D \leq 50$~kpc indicates
that $W_r(2796)$ is also not correlated with color at smaller impact
parameters ($0.5~\sigma$).

The lack of a correlation between $W_r(2796)$ and $B-K$ is consistent
with the findings of \citet{chen10a}, who examined $B-R$ colors for
$0.1 \leq z \leq 0.5$. However, the result is contrary to the
statistically based results of \citet{zibetti07}, who find bluer
colors from the integrated light selected by stronger {\MgII}
absorbers and redder colors selected by weaker absorbers for $0.4 \leq
z \leq 1$, where the minimum $W_r(2796)$ of their sample is
0.8~{\AA}. A direct comparison with the work of \citet{zibetti07} is
difficult due to its statistical nature. However, a Kendall's $\tau$
test examining the $W_r(2796)\geq 0.8$~{\AA} subsample of {\magiicat}
yielded a slight trend between $W_r(2796)$ and $B-K$ ($2.1~\sigma$)
such that larger equivalent widths may show a weak trend with galaxy
color. Bifurcating the full galaxy sample at $W_{\rm cut}=0.1$, 0.3,
0.6, and 1.0~{\AA}, we find a $2.5~\sigma$ significance for a
correlation between $W_r(2796)$ and $B-K$ for galaxies hosting
$W_r(2796) \geq 1.0$~{\AA} absorption; for the very strongest
absorbers, the redder the galaxy, the greater the {\MgII} equivalent
width. As mentioned, this is contrary to the findings of
\citet{zibetti07}.

To examine the possible {\MgII} equivalent width distribution
dependence on $B-K$ color, we also performed a Kolmogorov-Smirnov (KS)
test comparing the cumulative $W_r(2796)$ distribution functions of
red ($B-K \geq 1.48$) and blue ($B-K<1.48$) galaxies. We limited the
test to include only those galaxies with detected {\MgII}
absorption. The two subsamples are statistically consistent with
having been drawn from the same parent distribution ($\simeq
0.5~\sigma$). Limiting the sample to galaxies with $D \leq 50$~kpc,
the KS test results remain below $1~\sigma$. Though the distribution
for red galaxies has power beyond the largest and smallest $W_r(2796)$
of the blue galaxies, the statistics do not ferret out a difference
between the two galaxy subsamples; the average and maximum $W_r(2796)$
for $D \leq 50$~kpc associated with red galaxies and blue galaxies are
consistent (average 1.02 and 0.92~{\AA}, with maximum 2.9 and
2.3~{\AA}, respectively).

Our findings that the $W_r(2796)$ distributions from the CGM within
$D=50$~kpc of blue and red galaxies are indistinguishable is at odds
with the dramatic finding of \citet{bordoloi11}. They report a factor
of eight times larger equivalent width associated with blue galaxies
compared to red galaxies in stacked spectra for which the CGM is
probed within $D=50$~kpc.

\subsection{$W_r(2796)$ and Impact Parameter}
\label{sec:Wr-D}

\begin{figure}[th]
\includegraphics[angle=0,trim = 2mm 0mm 0mm 0mm,scale=0.73]{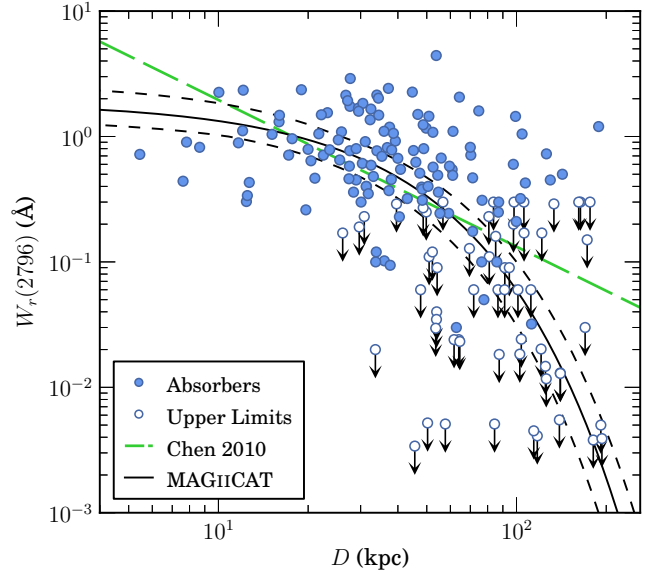}
\caption[]{The {\MgII}~$\lambda 2796$ rest-frame equivalent width,
  $W_r(2796)$, versus impact parameter, $D$. Galaxies with detected
  {\MgII} absorption are presented as solid blue points, whereas those
  with upper limits on absorption are open blue points with downward
  arrows. The full sample comprises 182 galaxies with a $7.9~\sigma$
  anti-correlation between $W_r(2796)$ and $D$. The green, long dashed
  curve is the power law fit obtained by \citet{chen10a} for their
  data. The solid curve is a log-linear maximum likelihood fit to the
  data presented here, $\log W_r(2796)=\alpha_1D+\alpha_2$, where
  $\alpha_1=-0.015 \pm 0.002$ and $\alpha_2=0.27 \pm 0.11$. Short
  dashed curves provide 1~$\sigma$ uncertainties in the fit.}
\label{fig:EWvsD}
\end{figure}

A commonly known property of {\MgII} galaxies is the anti-correlation
between $W_r(2796)$ and impact parameter, $D$,
\citep[e.g.,][]{lanzetta90, bb91, steidel95, bouche06, ggk08, chen10a,
  churchill-weakgals}. In Figure~\ref{fig:EWvsD}, we present
$W_r(2796)$ versus $D$. Galaxies with detected {\MgII} absorption are
plotted as solid blue points and those with upper limits on absorption
are plotted as open blue points with downward arrows.

We performed a non-parametric Kendall's $\tau$ rank correlation test
on $W_r(2796)$ against $D$, allowing for upper limits on
$W_r(2796)$. For this sample, $W_r(2796)$ is anti-correlated with $D$
at the $7.9~\sigma$ level. Since $W_r(2796)$ correlates with the
number of clouds \citep[Voigt profile
  components,][]{pb90,cvc03,evans-thesis}, this result indicates that
either the column densities, velocity spreads, or both, diminish with
projected distance from the galaxy.

To parameterize the general behavior of the anti-correlation, we used
the Expectation-Maximization maximum-likelihood method
\citep{wolynetz79} accounting for upper limits on $W_r(2796)$ to fit a
power law, $\log W_r(2796)=\alpha_1 \log D+\alpha_2$, and a log-linear
fit, $\log W_r(2796)=\alpha_1D+\alpha_2$, to the data. The data are
poorly described by a power law [green, long dashed curve in
  Figure~\ref{fig:EWvsD}, from \citet{chen10a}] due to the substantial
population of $W_r(2796) < 0.1$~{\AA} absorbers and upper limits. We
present the log-linear fit and its $1~\sigma$ uncertainties in
Figure~\ref{fig:EWvsD} (solid black curve) for $\alpha_1=-0.015 \pm
0.002$ and $\alpha_2=0.27 \pm 0.11$.

The considerable scatter about this relation {\it may\/} suggest that
$W_r(2796)$ is governed by physical processes related to the galaxy
such as luminosity \citep[cf.,][]{ggk08, chen10a}, mass
\citep[cf.,][]{bouche06, gauthier09, lundgren09, churchill-masses},
star formation \citep[cf.,][]{chen10b, menard11}, or orientation
\citep[cf.,][]{bordoloi11, bouche11, kacprzak, kcn,
  churchill-weakgals}.  Alternatively, the CGM is inherently patchy
\citep{cwc1317, churchill-weakgals}.

\begin{figure*}[bht]
\centering
\includegraphics[scale=0.8]{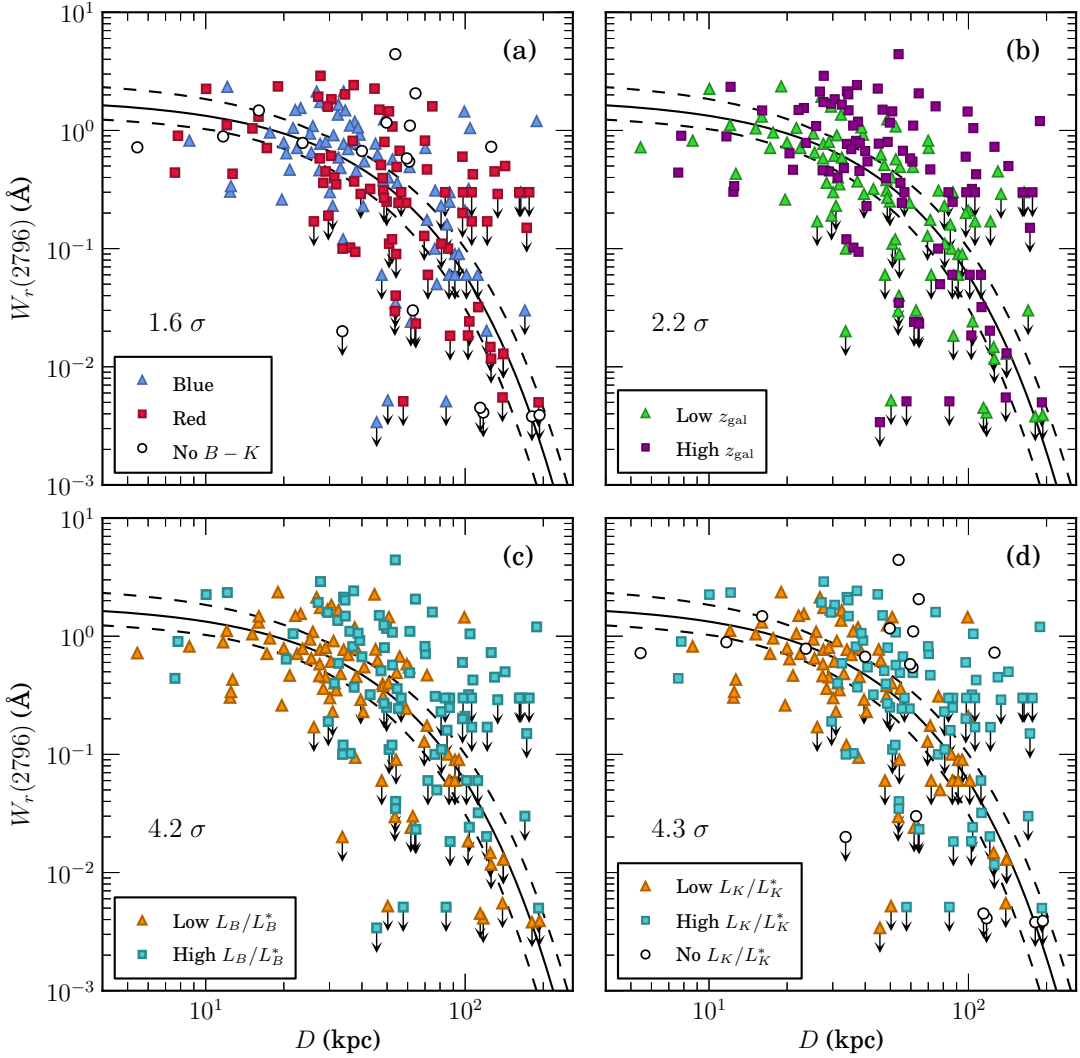}
\caption[]{The {\MgII} equivalent width, $W_r(2796)$, as a function of
  impact parameter, $D$, split by median ($a$) galaxy color, ($b$)
  redshift, ($c$) $B$-band luminosity, and ($d$) $K$-band
  luminosity. Galaxies with no $K$-band magnitude are presented as
  open points in panels $a$ and $d$. The null hypothesis that two
  subsamples are drawn from the same population can be ruled out to
  the $1.6~\sigma$ ($B-K$), $2.2~\sigma$ ($z_{\rm gal}$), $4.2~\sigma$
  ($L_B/L_B^{\ast}$), and $4.3~\sigma$ ($L_K/L_K^{\ast}$) level. The
  significance level decreases in all cases when we only consider
  galaxies for which we have detected {\MgII} absorption.}
\label{fig:EWD2cuts}
\end{figure*}

To examine possible trends and/or sources of this scatter in the
distribution of $W_r(2796)$ versus $D$, we applied the two-dimensional
Kolmogorov-Smirnov (2DKS) test. This test examines if the 2D
distribution of $W_r(2796)$ and $D$ from two samples can be ruled out
as being consistent.

We first investigated bifurcations of the full sample about the median
values of $B-K$, $z_{\rm gal}$, $L_B/L_B^{\ast}$, and
$L_K/L_K^{\ast}$, which are presented in
Figure~\ref{fig:EWD2cuts}. Performing a 2DKS test on the subsamples,
we find that the null hypothesis that blue and red galaxies are drawn
from the same population can not be ruled out to the $1.6~\sigma$
level. Using a redshift cut, the data show a weak trend for larger
$W_r(2796)$ at fixed $D$ for high $z$ galaxies ($2.2~\sigma$). A 2DKS
test on the luminosities shows that high $L_B/L_B^{\ast}$ and
$L_K/L_K^{\ast}$ are found with larger $W_r(2796)$ at fixed $D$ than
low luminosities at the $4.2~\sigma$ and $4.3~\sigma$ level,
respectively. We then examined only galaxies with detected absorption
in each subsample. In this case, the significance level decreases for
all subsamples.

We also divided the sample into quartiles by $B-K$, $z_{\rm gal}$,
$L_B/L_B^{\ast}$, and $L_K/L_K^{\ast}$ and performed the 2DKS test on
the lowest and highest quartiles. We obtained $P(\hbox{\sc ks})=0.03$,
$0.30$, $0.0007$, and $0.0006$ for $B-K$, $z_{\rm gal}$,
$L_B/L_B^{\ast}$, and $L_K/L_K^{\ast}$, respectively. At best, there
is a $3.4~\sigma$ significance that the distribution or scatter of
$W_r(2796)$ versus $D$ is due to a dependence on $L_K/L_K^{\ast}$, a
$3.4~\sigma$ significance of a connection to $L_B/L_B^{\ast}$, and a
$2.1~\sigma$ significance it is connected to $B-K$. There is clearly
no redshift dependence. When we examine only galaxies for which we
have detectable absorption, we find that the significance level
decreases or remains the same for all subsamples.

\begin{figure*}[bht]
\centering
\includegraphics[scale=0.58]{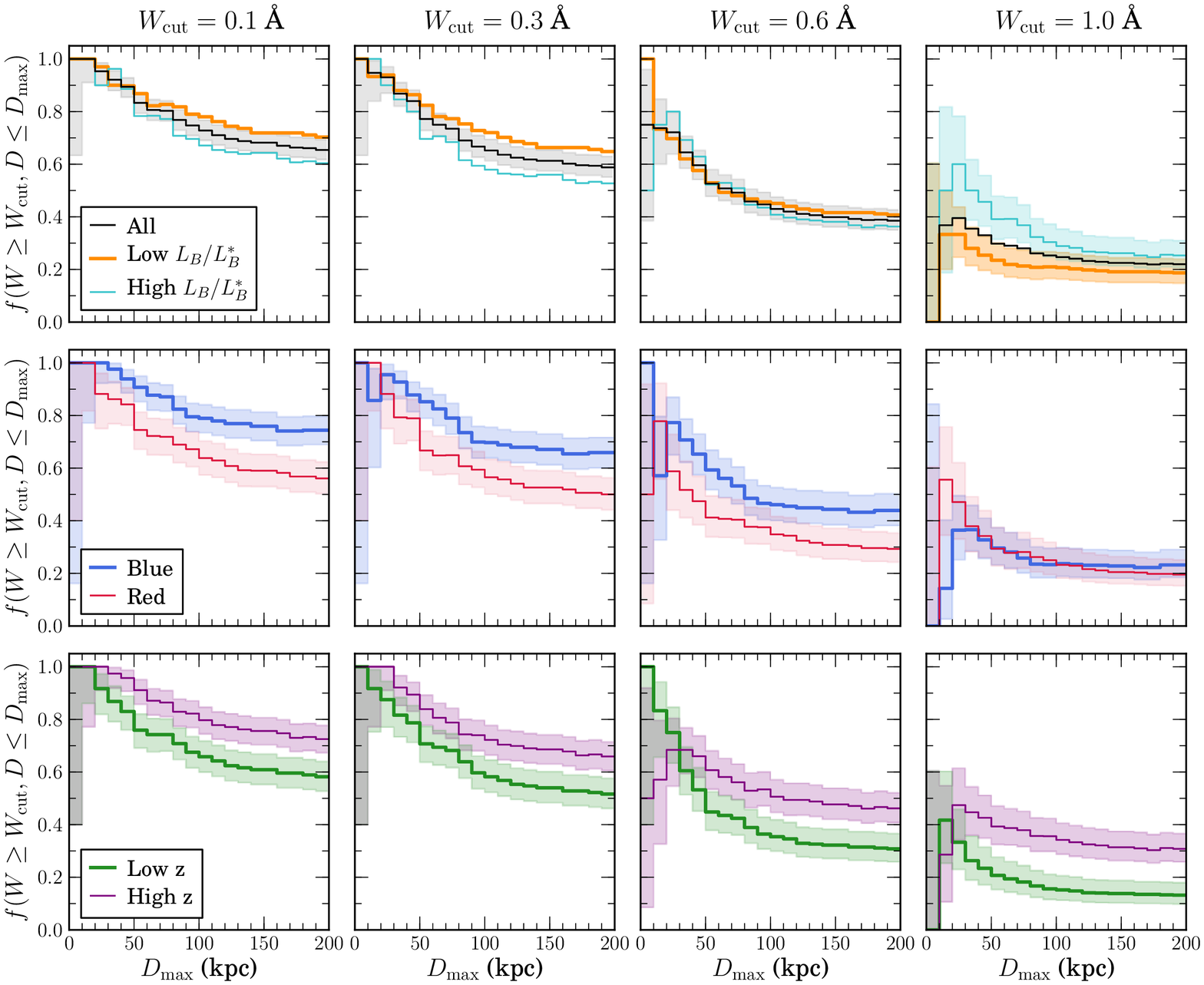}
\caption[]{The covering fraction, {\fdmax}, inside $D_{\rm max}$ for
  different bifurcated subsamples split at the median
  $L_B/L_B^{\ast}$, $B-K$, and $z_{\rm gal}$ and for various
  $W_r(2796)>W_{\rm cut}$ thresholds. Shaded regions represent
  $1~\sigma$ uncertainties based upon binomial statistics. --- (upper)
  The full sample of galaxies (thin black line), the high luminosity
  galaxies (thin blue line), and the low luminosity galaxies (thick
  orange line), divided at the median $B$-band luminosity,
  $L_B/L_B^{\ast}=0.611$. --- (middle) The blue ($B-K<1.48$) and red
  ($B-K \geq 1.48$) galaxy subsamples bifurcated at the median galaxy
  color. --- (lower) The high ($z \geq \langle z \rangle$) and low ($z
  < \langle z \rangle$) redshift galaxy subsamples where $\langle z
  \rangle = 0.359$.}
\label{fig:fdmax}
\end{figure*}

\begin{figure*}[thb]
\centering
\includegraphics[scale=0.58]{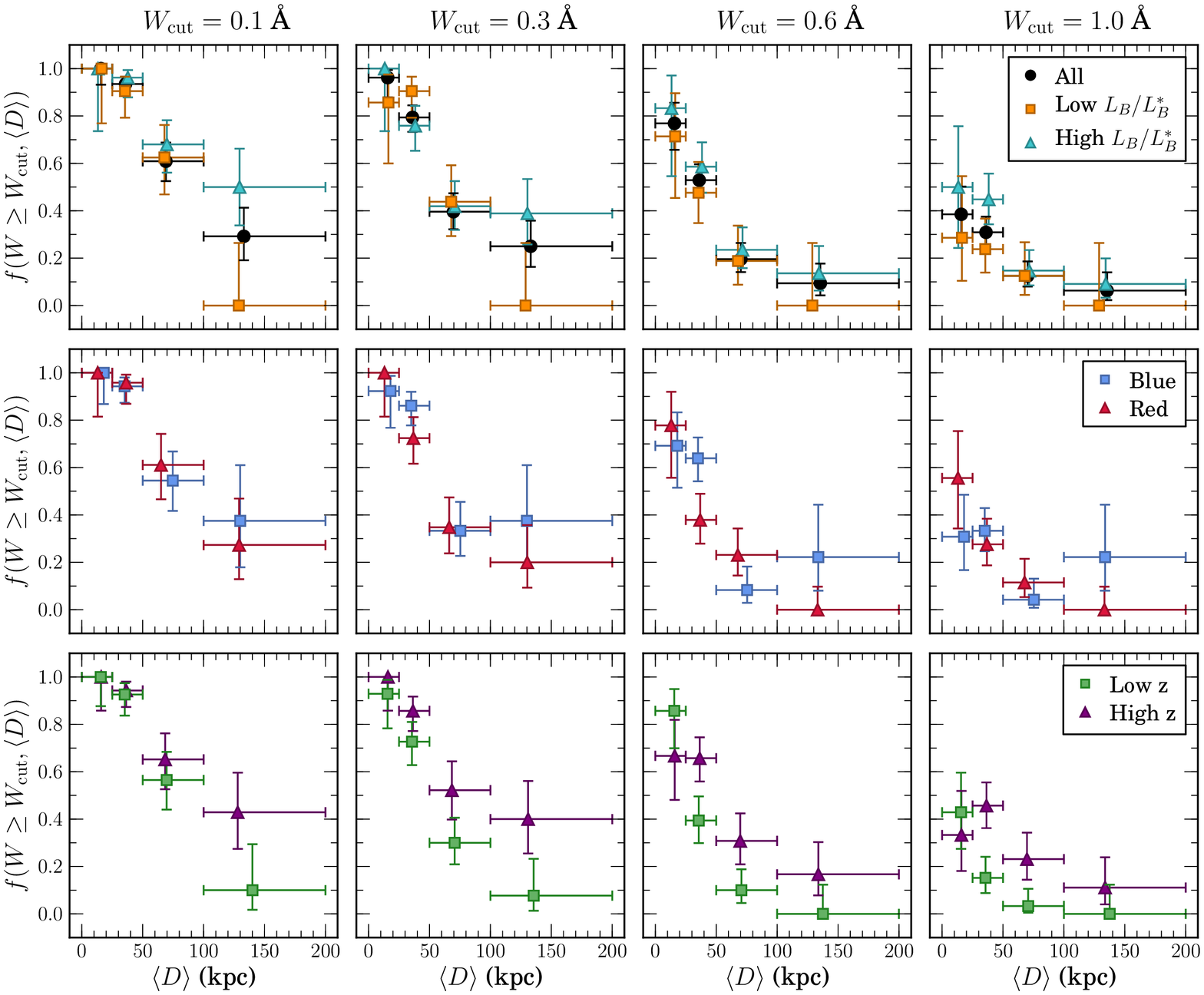}
\caption[]{The covering fraction profile, {\fdbar}, for the impact
  parameter bins $0 \leq D<25$~kpc, $25 \leq D<50$~kpc, $50 \leq
  D<100$~kpc, and $100 \leq D<200$~kpc for different $W_r(2796)$
  thresholds, $W_{\rm cut}$. The horizontal bars indicate the impact
  parameter bin width and the vertical bars are the $1~\sigma$
  binomial uncertainties. Data points are plotted at the mean impact
  parameter of the galaxies in the bin. --- (upper) The full sample of
  galaxies (solid black circles), the high luminosity galaxies (blue
  triangles), and the low luminosity galaxies (orange squares). ---
  (middle) Blue (blue squares) and red (red triangles) galaxy
  subsamples sliced at the median galaxy color. -- (lower) High
  (purple triangles) and low (green squares) redshift subsamples cut
  by the median redshift, $\langle z \rangle = 0.359$.}
\label{fig:fdbar}
\end{figure*}

\subsection{Covering Fraction and Impact Parameter}
\label{sec:fcov-D}

Here we examined the dependence of covering fraction directly on
impact parameter and galaxy luminosity, color, and redshift. We
present directly observable quantities, making no assumptions with
regard to the {\MgII} absorbing CGM density profile or geometry.

\subsubsection{Covering Fraction Within Fixed Impact Parameters}

In order to examine the covering fraction within a fixed projected
radial distance from the galaxies, we computed the quantity ${\fdmax}
\equiv f(W \geq W_{\rm cut},D \leq D_{\rm max})$, which we define as
the fraction of absorbers with $W_r(2796) \geq W_{\rm cut}$ inside a
projected separation of $D_{\rm max}$ from the galaxy.

In Figure~\ref{fig:fdmax}, we plot {\fdmax} against $D_{\rm max}$ in
10~kpc intervals. The uncertainties in {\fdmax} are shown as shaded
regions. These are the upper and lower limits, calculated using the
formalism for binomial statistics \citep[see][]{gehrels86}. Values of
{\fdmax} at $D_{\rm max} \leq 10$~kpc are not robust given the small
number of galaxies within this impact parameter\footnote{Of the three
  galaxies at $D \leq 10$~kpc, one is at low redshift where the
  angular separation from the quasar is much greater than the quasar
  seeing disk, and the other two were found at higher redshift
  following point spread function subtraction of the quasar
  \citep{sdp94}, which was not performed in all surveys.}. Due to the
relative undersampling of galaxies at $D_{\rm max}>150$~kpc and the
cumulative nature of {\fdmax}, the covering fraction will be virtually
unchanging outside this impact parameter; thus the shape of {\fdmax}
versus $D_{\rm max}$ is less robust for $D_{\rm max} > 150$~kpc.

In Figure~\ref{fig:fdmax} (upper panels), we present {\fdmax} for the
full galaxy sample and for low and high luminosity galaxies. For each
$W_{\rm cut}$ subsample, the covering fraction of the {\MgII}
absorbing CGM inside $D=D_{\rm max}$ decreases as $D_{\rm max}$ is
increased. At a given $D_{\rm max}$, there is a clear trend that as
$W_{\rm cut}$ is increased, {\fdmax} decreases, indicating that the
covering fraction within a fixed projected separation increases as the
minimum absorption threshold is lowered. The covering fraction may
exhibit a luminosity dependence for $W_{\rm cut}=0.1$ and 0.3~{\AA}
such that low luminosity galaxies have larger covering fractions than
high luminosity galaxies, but this is most apparent outside of
$D=100$~kpc where the covering fraction is less robust. We find no
luminosity dependence for $W_{\rm cut} = 0.6$~{\AA}. For $W_{\rm
  cut}=1.0$~{\AA}, {\fdmax} is larger for high luminosity galaxies at
all $D_{\rm max}$ than for low luminosity galaxies. This difference is
accentuated at $D_{\rm max} \leq 50$~kpc.

In the middle panels of Figure~\ref{fig:fdmax}, we present {\fdmax}
and its uncertainties versus $D_{\rm max}$ for red and blue
galaxies. The mean redshift for red galaxies is $\overline{z}=0.36$
over the range $0.1 \leq z \leq 0.85$, and for blue galaxies
$\overline{z}=0.44$ over $0.1 \leq z \leq 1.02$. For the $W_{\rm cut}
= 0.1$, 0.3, and 0.6~{\AA} subsamples, blue galaxies have higher
covering fractions relative to red galaxies for $D_{\rm
  max}>50$~kpc. Within 50~kpc, the larger uncertainties make it
difficult to distinguish any possible differences. For $W_{\rm
  cut}=1.0$~{\AA}, the covering fractions for red and blue galaxies
are indistinguishable outside of $D=50$~kpc. Within $D=50$~kpc, red
galaxies may have larger covering fractions, though the uncertainties
are again large.

In Figure~\ref{fig:fdmax} (lower panels), we present {\fdmax} and its
uncertainties against $D_{\rm max}$ for the high $z$ and low $z$
galaxy subsamples. For $D_{\rm max} \leq 50$~kpc, it is difficult to
distinguish any redshift evolution in {\fdmax} due to the
uncertainties. Beyond 50~kpc, the data suggest that {\fdmax} may
evolve such that at higher redshift, the covering fraction of {\MgII}
absorbing gas is higher than at lower redshift. 

\citet{chen10a} also examined the luminosity dependence of ${\fdmax}$
as a function of $W_{\rm cut}$, where their $D_{\max}$ is fixed at the
luminosity scaled ``gas radius'', with $R_{\rm gas } \propto
(L_B/L_B^{\ast})^{0.35}$, assuming isothermal density profile
\citep{tinkerchen08} and NFW profile \citep{nfw96} models. As such, a
direct comparison with our non-parameterized results is
difficult. Nonetheless, they found that ${\fdmax}$ has little to no
dependence on galaxy $B$-band luminosity for $W_{\rm cut}=0.1$, $0.3$,
and $0.5$~{\AA}, which is consistent with our result; however, for
$W_{\rm cut}=1.0$~{\AA}, we found systematically larger ${\fdmax}$ for
high luminosity galaxies as compared to low luminosity galaxies,
especially for $D_{\rm max} \leq 50$~kpc.

\subsubsection{Covering Fraction Profiles}

To examine the covering fraction profile with projected distance from
the galaxy, we computed ${\fdbar} \equiv f(W \geq W_{\rm cut},\langle
D \rangle)$, which we define as the fraction of absorbers with
$W_r(2796) \geq W_{\rm cut}$ in fixed impact parameter bins. We select
the bins $0 \leq D<25$~kpc, $25 \leq D<50$~kpc, $50 \leq D<100$~kpc,
and $100 \leq D<200$~kpc, for which $\langle D \rangle$ is the average
impact parameter of the galaxies in the bin.

In Figure~\ref{fig:fdbar}, we plot {\fdbar} against $\langle D
\rangle$. The horizontal bars are the impact parameter ranges and the
data points are the average impact parameters, $\langle D \rangle$,
for each bin. The vertical error bars are the $1~\sigma$ uncertainties
in {\fdbar} based upon binomial statistics. The subsamples presented
in each panel are identical to the corresponding panels of
Figure~\ref{fig:fdmax}.

As can be seen in the upper panels of Figure~\ref{fig:fdbar}, the
{\fdbar} profile decreases as impact parameter is increased and this
behavior is exhibited regardless of $W_{\rm cut}$. Also note that as
$W_{\rm cut}$ is increased, {\fdbar} is smaller for a given impact
parameter bin. This indicates that in a fixed annulus around galaxies,
the sky-projected distribution of the {\MgII} CGM becomes
progressively patchier as higher column density material and/or more
complex kinematics are selected. For $W_{\rm cut}=0.1$~{\AA}, the
{\fdbar} profile decreases from unity within a radius of 25~kpc to
30\% in the annulus 100--200~kpc, whereas for $W_{\rm cut}=1.0$~{\AA},
the covering fraction decreases from 40\% to less than 10\%.

\begin{deluxetable}{ccccc}
\tablecolumns{5}
\tablewidth{0pt}
\tablecaption{Luminosity Dependence of Covering Fraction Profiles, ${\fdbar}$ \label{tab:fdbar}}
\tablehead{
\colhead{$W_{\rm cut}$, {\AA}} &
\colhead{(0--25), kpc}      &
\colhead{[25--50), kpc}     &
\colhead{[50--100), kpc}    &
\colhead{[100--200), kpc}   
}
\startdata
\cutinhead{All Galaxies} 
\\[-7pt]
0.1   & $1.00_{-0.07}^{+0.00}$ & $0.94_{-0.05}^{+0.03}$ & $0.61_{-0.08}^{+0.08}$ & $0.29_{-0.10}^{+0.12}$ \\[3pt]
0.3   & $0.96_{-0.08}^{+0.03}$ & $0.79_{-0.06}^{+0.05}$ & $0.40_{-0.07}^{+0.08}$ & $0.25_{-0.09}^{+0.11}$ \\[3pt]  
0.6   & $0.77_{-0.11}^{+0.09}$ & $0.53_{-0.07}^{+0.07}$ & $0.20_{-0.06}^{+0.07}$ & $0.09_{-0.05}^{+0.08}$ \\[3pt]  
1.0   & $0.39_{-0.11}^{+0.12}$ & $0.31_{-0.06}^{+0.07}$ & $0.13_{-0.05}^{+0.06}$ & $0.06_{-0.04}^{+0.08}$ \\[-2pt]
\cutinhead{Low $L_B/L_B^{\ast}$ Galaxies}
\\[-7pt]
0.1   & $1.00_{-0.23}^{+0.00}$ & $0.91_{-0.11}^{+0.06}$ & $0.63_{-0.16}^{+0.14}$ & $0.00_{-0.00}^{+0.26}$ \\[3pt]
0.3   & $0.86_{-0.26}^{+0.12}$ & $0.91_{-0.11}^{+0.06}$ & $0.44_{-0.15}^{+0.15}$ & $0.00_{-0.00}^{+0.26}$ \\[3pt]   
0.6   & $0.71_{-0.26}^{+0.18}$ & $0.48_{-0.13}^{+0.13}$ & $0.19_{-0.10}^{+0.15}$ & $0.00_{-0.00}^{+0.26}$ \\[3pt]   
1.0   & $0.29_{-0.18}^{+0.26}$ & $0.24_{-0.10}^{+0.13}$ & $0.13_{-0.08}^{+0.14}$ & $0.00_{-0.00}^{+0.26}$ \\[-2pt]   
\cutinhead{High $L_B/L_B^{\ast}$ Galaxies}
\\[-7pt]
0.1   & $1.00_{-0.26}^{+0.00}$ & $0.96_{-0.08}^{+0.03}$ & $0.68_{-0.12}^{+0.10}$ & $0.50_{-0.16}^{+0.16}$ \\[3pt]
0.3   & $1.00_{-0.26}^{+0.00}$ & $0.76_{-0.10}^{+0.08}$ & $0.42_{-0.10}^{+0.11}$ & $0.39_{-0.13}^{+0.15}$ \\[3pt] 
0.6   & $0.83_{-0.29}^{+0.14}$ & $0.59_{-0.11}^{+0.10}$ & $0.24_{-0.08}^{+0.10}$ & $0.14_{-0.07}^{+0.12}$ \\[3pt] 
1.0   & $0.50_{-0.26}^{+0.26}$ & $0.45_{-0.11}^{+0.11}$ & $0.15_{-0.06}^{+0.09}$ & $0.09_{-0.06}^{+0.11}$ \\[-2pt]
\enddata
\end{deluxetable}

Also shown in the upper panels of Figure~\ref{fig:fdbar} is {\fdbar}
split by $L_B/L_B^{\ast}=0.611$. For low luminosity galaxies, the
covering fraction vanishes outside of $D=100$~kpc for all $W_{\rm
  cut}$. On the other hand, in the annulus 100--200~kpc, the covering
fractions for high luminosity galaxies are approximately 50\%, 40\%,
15\%, and 10\% for $W_{\rm cut}=0.1$, 0.3, 0.6, and 1.0~{\AA},
respectively. Covering fraction profiles for all, low, and, high
luminosity galaxies are listed in Table~\ref{tab:fdbar}.

The data are suggestive of a luminosity dependence for the maximum
extent of {\MgII} absorbing gas, such that the CGM of low
$L_B/L_B^{\ast}$ galaxies extends no further than 100~kpc, whereas
high luminosity galaxies have a detectable {\MgII} CGM beyond
100~kpc. The difference between high and low luminosity galaxies is
most prominent outside of 100~kpc for $W_{\rm cut}=0.1$, and
0.3~{\AA}. As such, high luminosity galaxies exhibit a more extended
{\MgII} CGM, whereas for low luminosity galaxies, the {\MgII} CGM is
more concentrated in the central regions but with relatively lower
covering fraction as $W_{\rm cut}$ is increased.

In the middle panels of Figure~\ref{fig:fdbar}, the covering fraction
profiles of red and blue galaxies exhibit several differences. Note
that for red galaxies, none have absorption with $W_r(2796) \geq
0.6$~{\AA} in the range 100--200~kpc. Blue galaxies consistently have
covering fractions of approximately 40\% for $W_{\rm cut} \leq
0.3$~{\AA} and 20\% for $W_r(2796) \geq 0.6$~{\AA} in the range
100--200~kpc. Within 25~kpc the covering fraction for blue galaxies
drops more rapidly than for red galaxies with increasing $W_{\rm
  cut}$. Moreover, for $W_{\rm cut}=0.6$ and $1.0$~{\AA} there is a
hint that the covering fraction of red galaxies drops more rapidly for
$D>25$ kpc relative to the $D=0$--25 kpc region. This may suggest that
the gas in red galaxies is more concentrated near the center while in
blue galaxies, the gas is not as highly concentrated.

\begin{figure*}[bth]
\centering
\includegraphics[scale=0.59]{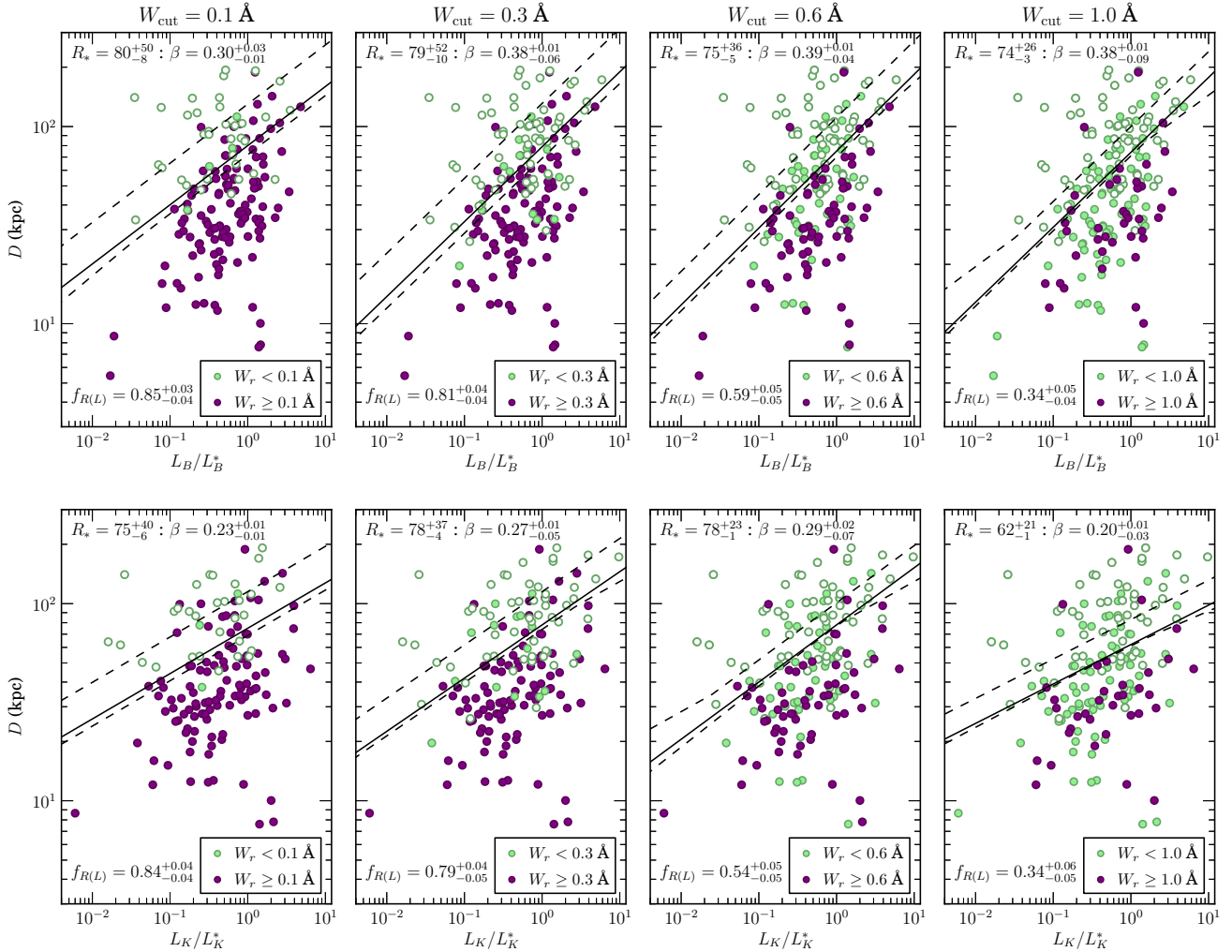}
\caption[]{Impact parameter, $D$, versus luminosity for different
  $W_r(2796)$ bifurcations at $W_{\rm cut}$. Green points are galaxies
  with $W_r(2796) < W_{\rm cut}$ (solid points) or an upper limit on
  absorption (open points) and purple points are galaxies with
  $W_r(2796) \geq W_{\rm cut}$. --- (upper) $B$-band luminosity. ---
  (lower) $K$-band luminosity. The fit parameters $R_{\ast}$, $\beta$,
  and {\frl} give the absorbing gas halo radius for an $L^{\ast}$
  galaxy, luminosity scaling power index, and absorption covering
  fraction, respectively, for each $W_r(2796)=W_{\rm cut}$ threshold
  (see Table~\ref{tab:halos}). Dashed lines provide the envelope
  1~$\sigma$ uncertainties in the fit at fixed $L/L^{\ast}$. For
  $W_{\rm cut}=0.1$~{\AA}, galaxies with an upper limit on absorption
  greater than 0.1~{\AA} were not included in the fitting process and
  are therefore not plotted. }
\label{fig:DLenv}
\end{figure*}

As shown in the lower panels of Figure~\ref{fig:fdbar}, high redshift
galaxies exhibit a trend of higher {\fdbar} in the 25--50, 50--100,
and 100--200~kpc annuli relative to low redshift galaxies. Within
25~kpc, the covering fraction may be higher in low redshift galaxies
for stronger absorption, $W_{\rm cut}=0.6$ and 1.0~{\AA}. Though the
trends are marginal, they may suggest a greater extension of the gas
at higher redshift followed by a settling of material into the inner
regions at lower redshift. 

\citet{chen10a} examined ${\fdbar}$ for their sample ($\langle z
\rangle = 0.25$), and also found that the covering fraction profile,
{\fdbar}, decreases with increasing $D$ and is smaller at a given
$\langle D \rangle$ with increasing $W_{\rm cut}$ using $W_{\rm
cut}=0.1$, $0.3$, and $0.5$~{\AA}. However, a direct comparison with
our results is difficult because \citet{chen10a} scale $D$ assuming a
$B$-band galaxy luminosity dependent ``gas radius'' proportional to
$(L_B/L_B^{\ast})^{0.35}$.

\subsection{Luminosity Scaling and Covering Fraction}
\label{sec:RL-fcov}

Our results in the previous section strongly suggest that the {\MgII}
absorbing CGM extends further for weaker absorption than for stronger
absorption, and that this behavior has a clear dependence with galaxy
luminosity. What is further clear, is that for a given $W_{\rm cut}$,
the CGM becomes more ``patchy'' with increasing $D$, as indicated by
the decreasing covering fractions. {\it As such, it would seem that
  the notion of a well-defined {\MgII} CGM ``gas radius'', or ``halo
  absorption radius'' is an oversimplification in that the extent of
  the gas exhibits a ``fuzzy'' boundary when averaged over many
  galaxies}. Nonetheless, for historical comparisons we parameterize
the characteristics of an ``outer'' boundary and its plausible
dependence on the galaxy properties and redshift.

\begin{deluxetable}{ccccccc}
\tablecolumns{7}
\tablewidth{0pt}
\tablecaption{Luminosity Scaled Halo Absorption Radii \label{tab:halos}}
\tablehead{
\colhead{\phantom{x}}       &
\multicolumn{3}{c}{------------- $B$-band -------------} &
\multicolumn{3}{c}{------------- $K$-band -------------} \\
\colhead{$W_{\rm cut}$} & 
\colhead{$R_{\ast}$}          & 
\colhead{$\beta$}            &  
\colhead{{\frl}}        &  
\colhead{$R_{\ast}$}          & 
\colhead{$\beta$}            &  
\colhead{{\frl}}        \\
\colhead{[{\AA}]} & 
\colhead{[kpc]}          & 
\colhead{}            &  
\colhead{}        &  
\colhead{[kpc]}          & 
\colhead{}            &  
\colhead{}        
}
\startdata
0.1 & $80^{+50}_{-8}$ & $0.30^{+0.03}_{-0.01}$ & $0.85^{+0.03}_{-0.04}$ & $75^{+40}_{-6}$ & $0.23^{+0.01}_{-0.01}$ & $0.84^{+0.04}_{-0.04}$ \\[3pt]
0.3 & $79^{+52}_{-10}$ & $0.38^{+0.01}_{-0.06}$ & $0.81^{+0.04}_{-0.04}$ & $78^{+37}_{-4}$ & $0.27^{+0.01}_{-0.05}$ & $0.79^{+0.04}_{-0.05}$ \\[3pt]
0.6 & $75^{+36}_{-5}$ & $0.39^{+0.01}_{-0.04}$ & $0.59^{+0.05}_{-0.05}$ & $78^{+23}_{-1}$ & $0.29^{+0.02}_{-0.07}$ & $0.54^{+0.05}_{-0.05}$ \\[3pt]
1.0 & $74^{+26}_{-3}$ & $0.38^{+0.01}_{-0.09}$ & $0.34^{+0.05}_{-0.04}$ & $62^{+21}_{-1}$ & $0.20^{+0.01}_{-0.03}$ & $0.34^{+0.06}_{-0.05}$ \\[-5pt]
\enddata
\end{deluxetable}

Since the work of \citet{bb91}, the extent of absorbing gas is
commonly assumed to follow a Holmberg-like relation,
$R(L)=R_{\ast}(L/L^{\ast})^{\beta}$. We examined whether the halo
absorption radius also depends on $W_r(2796)$, and/or galaxy color and
redshift. 

To examine the dependence on $W_r(2796)$, we adopt $W_{\rm cut}=0.1$,
0.3, 0.6, and 1.0~{\AA}. To estimate $R_{\ast}$ and $\beta$, we varied
the two parameters over the ranges $0 \leq R_{\ast} \leq 300$~kpc and
$0 \leq \beta \leq 1$ and computed the function
$q(R_{\ast},\beta,L,W_{\rm cut}) = wr_{(\geq)} + r_{(<)}$, where
$r_{(\geq)}$ is the fraction of systems with $W_r(2796) \geq W_{\rm
  cut}$ below $R(L)$, $r_{(<)}$ is the fraction with $W_r(2796) <
W_{\rm cut}$ above $R(L)$, and $w$ is a weighting
factor\footnote{Since we wish to determine the ``outer envelope'' of
  the halo absorption radius, we adopt the weight $w=2$. When $w=1$,
  we find $\beta$ consistent with zero.}. The parameter values are
adopted when $q(R_{\ast},\beta,L,W_{\rm cut})$ is a maximum, for which
the covering fraction inside $R(L)$ is ${\frl} = 1/(1+x)$, where
$x=(n_{(\geq)}/n_{(<)})(1-r_{(<)})/r_{(\geq)}$, and where $n_{(\geq)}$
and $n_{(<)}$ are the number of systems with $W_r(2796) \geq W_{\rm
  cut}$ and $W_r(2796) < W_{\rm cut}$, respectively.

The downward and upward $1~\sigma$ uncertainties in $R_{\ast}$ were
estimated by using the best estimate of $\beta$ and performing
one-sided integration under the $q(R_{\ast},\beta,L,W_{\rm cut})$
curve until 84.13\% of the total area was obtained. Similarly, the
downward and upward $1~\sigma$ uncertainties in $\beta$ were estimated
by using the best estimate of $R_{\ast}$. The $1~\sigma$
uncertainties in {\frl} are computed using binomial statistics by
computing the fraction of galaxies with $W_r(2796) \geq W_{\rm cut}$
to all galaxies within $R(L)$.

Our results are presented in Table~\ref{tab:halos} for both $B$- and
$K$-band luminosities. In the upper panels of Figure~\ref{fig:DLenv},
we present $D$ against $L_B/L^{\ast}_B$. The luminosity scaling
increases from $\beta \sim 0.3$ for $W_{\rm cut}=0.1$~{\AA} to $\beta
\sim 0.4$ for $W_{\rm cut}=0.3$, 0.6, and 1.0~{\AA}. The absorbing gas
halo radius, $R_{\ast}$, for an $L_B^{\ast}$ galaxy is on the order of
$\sim 80$~kpc for $W_{\rm cut}=0.1$ and 0.3~{\AA}, while $R_{\ast}$ is
on the order of $\sim 75$~kpc (but consistent with $80$~kpc, given the
uncertainties in $R_{\ast}$) for $W_{\rm cut}=0.6$ and 1.0~{\AA}. The
covering fraction decreases from ${\frl}=0.85$ for $W_{\rm
  cut}=0.1$~{\AA} to ${\frl}=0.34$ for $W_{\rm cut}=1.0$~{\AA}.

In the lower panels of Figure~\ref{fig:DLenv}, we present the $K$-band
results. We note that this subsample has 18 fewer galaxies than the
$B$-band subsample. The luminosity scaling varies from $\sim 0.2$ for
$W_{\rm cut}=0.1$ and 1.0~{\AA} to $\sim 0.3$ for $W_{\rm cut}=0.3$
and 0.6~{\AA}. The halo absorption radius also has a range of values
where $R_{\ast}=75$~kpc for $W_{\rm cut}=0.1$~{\AA}, 78~kpc for 0.3
and 0.6~{\AA}, and 62~kpc for 1.0~{\AA}, however, these values are all
consistent within uncertainties. The covering fraction behaves
similarly to the $B$-band (see Table~\ref{tab:halos}).

\citet{steidel95} reported $\beta \simeq 0.2$ and $R_{\ast}=55$~kpc
for the $B$-band and $\beta \simeq 0.15$ and $R_{\ast}=58$~kpc for the
$K$-band, both for $W_{\rm cut}=0.3$~{\AA} (the values of $R_{\ast}$
quoted here have been converted to a ``737'' $\Lambda$CDM cosmology at
the mean redshift, $\langle z \rangle=0.65$, of his sample). He
deduced a covering fraction of ${\frl} \simeq 1$ and that $\beta$ and
$R_{\ast}$ for the $K$-band are slightly smaller than for the
$B$-band. \citet{guillemin} obtained similar results, with
$R_{\ast}=47$~kpc (adjusted from $q_0=0.05$ and $h_{50}$ at $z=0.7$)
and $\beta=0.28$ for the $B$-band with $W_{\rm cut} \simeq 0.3$~{\AA}.

Using the observed {\MgII} absorber redshift path number density as a
constraint, \citet{ggk08} explored the parameter space of the minimum
luminosity of {\MgII} absorption-selected galaxies; the slope,
$\beta$; the halo absorption radius, $R_{\ast}$; and covering
fraction, {\frl}. For $W_{\rm cut}=0.3$~{\AA}, they constrained
$R_{\ast}$ to the range 50--110~kpc, $\beta$ to the range 0.2--0.28,
and {\frl} to the range 50--80\%. Our values of $R_{\ast}$ reside in
the range of allowed values from their study.

\begin{figure}[thb]
\includegraphics[clip=true,trim = 2mm 0mm 0mm 0mm,scale=0.57]{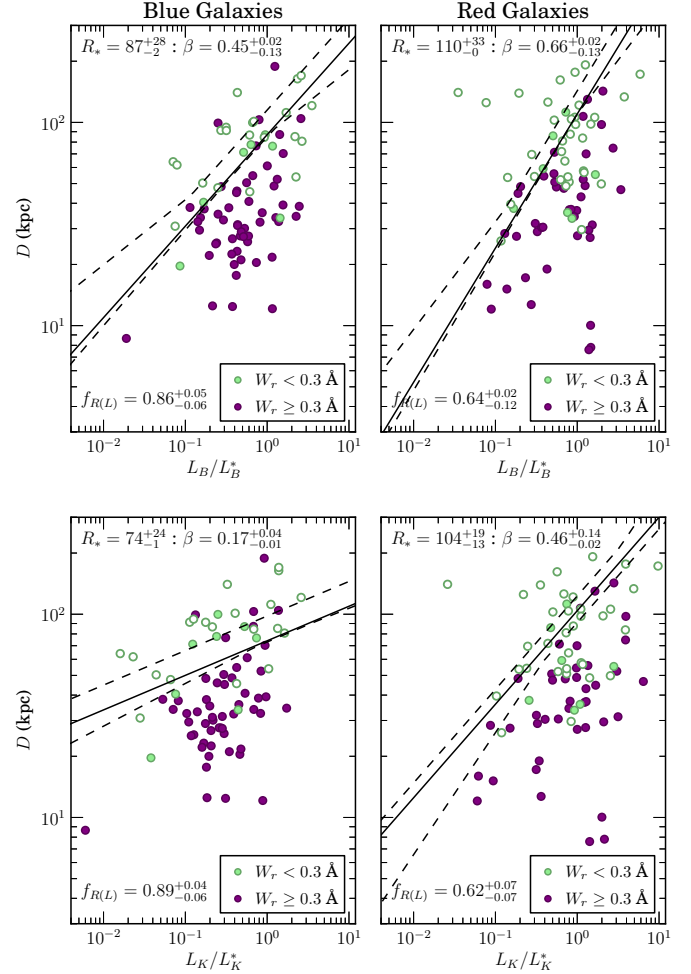}
\caption[]{Impact parameter, $D$, versus $L_B/L^{\ast}_B$ (top) and
  $L_K/L^{\ast}_K$ (bottom) for blue and red galaxies split by the
  median color $B-K=1.48$. Point types and curves are the same as
  Figure~\ref{fig:DLenv} for $W_{\rm cut} = 0.3$~{\AA}. The luminosity
  scaling, $\beta$, and radius, $R_{\ast}$, of the absorbing gas halo
  radius exhibits a color dependence for this $W_{\rm cut}$. In the
  $B$-band, blue galaxies have a steeper luminosity scaling, larger
  $R_{\ast}$, and larger covering fraction than red galaxies. For the
  $K$-band, we find $R_{\ast}$ and $\beta$ to be larger in red
  galaxies than blue, while blue galaxies still have a larger covering
  fraction.}
\label{fig:typeenv}
\end{figure}

\begin{figure}[thb]
\includegraphics[clip=true,trim = 2mm 0mm 0mm 0mm,scale=0.57]{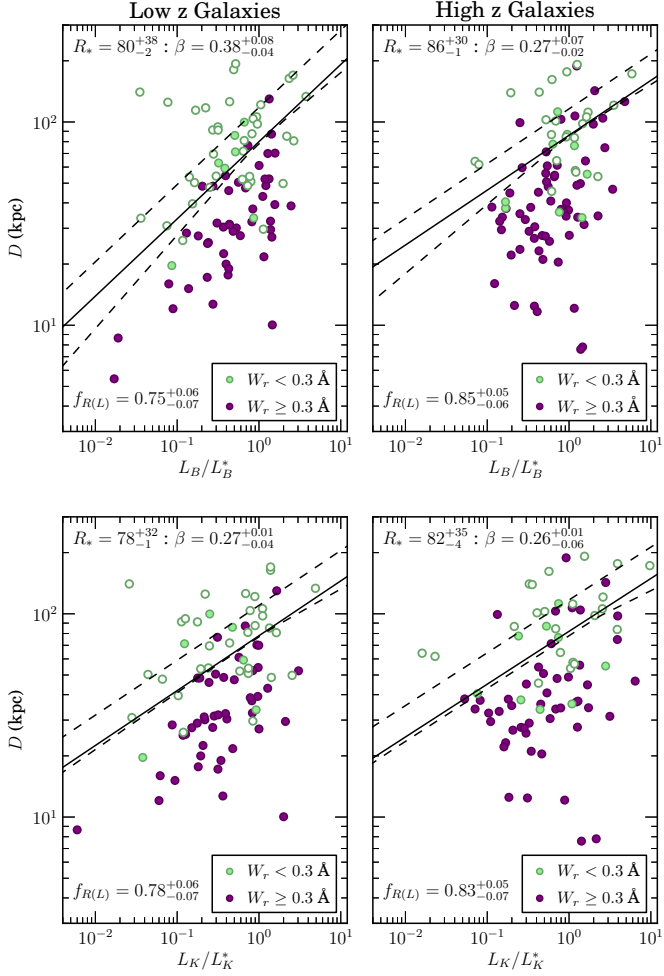}
\caption[]{Impact parameter, $D$, versus $L_B/L^{\ast}_B$ (top) and
  $L_K/L^{\ast}_K$ (bottom) for low and high redshift galaxies sliced
  by the median redshift $z_{\rm gal}=0.359$. Point types and curves
  are the same as Figure~\ref{fig:DLenv} for $W_{\rm cut} =
  0.3$~{\AA}. The average $z$ for the low redshift subsample is 0.23
  and the average for the high redshift subsample is 0.61, which is a
  $\sim 3.2$~Gyr difference.}
\label{fig:DLz}
\end{figure}

\citet{chen10a} reported $R_{\ast}=107$~kpc with $\beta=0.35$ for the
$B$-band, where they fitted their data in the context of an isothermal
sphere model \citep[see][]{tinkerchen08} of the {\MgII} CGM. For these
values, they find ${\frl}=0.7$ for $W_{\rm cut}=0.3$~{\AA} and
${\frl}=0.8$ for $W_{\rm cut}=0.1$~{\AA}, both of which are $\sim
10$\% smaller than what we find.

Employing the \citet{tinkerchen08} isothermal sphere model,
\citet{bordoloi11} deduced $R_{\ast}=115$~kpc. Whereas $\beta$ for
this work is consistent with the value reported by \citet{chen10a},
the larger values of $R_{\ast}$ reported by both \citet{chen10a} and
\citet{bordoloi11} may be an artifact of their application of a model
to the data. Furthermore, the methods of \citet{bordoloi11} involve
``averaging'' over annuli of fixed impact parameter, which may
introduce an additional systematic.

\subsubsection{Galaxy Color}

To examine the dependence of $R(L)$ on galaxy color, we computed
$R_{\ast}$, $\beta$, and {\frl} for red and blue galaxies, bifurcated
by $B-K=1.48$, in both the $B$- and $K$-bands. Due to the smaller
number of galaxies in the blue and red subsamples (82 each), we
adopted a single equivalent width cut, $W_{\rm cut}=0.3$~{\AA} (the
median $W_r(2796)$ for galaxies with measured colors). As shown in
Figure~\ref{fig:typeenv} and tabulated in Table~\ref{tab:BKzhalos},
the parameters for $R(L)$ suggest a dependence on galaxy color in both
the $B$- and $K$-bands.

In the $B$-band (top panels of Figure~\ref{fig:typeenv}), we find that
red galaxies have a steeper luminosity scaling, $\beta \sim 0.7$, than
blue galaxies, $\beta \sim 0.5$, but a smaller covering fraction,
${\frl}=0.6$, than blue galaxies, ${\frl}=0.9$. The halo absorption
radius for red galaxies, $R_{\ast} \sim 110$~kpc, is larger than for
blue galaxies, $R_{\ast} \sim 87$~kpc. However, both values of
$R_{\ast}$ are consistent within uncertainties. These values are
smaller than the model dependent values found by \citet{bordoloi11},
who report $118$~kpc for red galaxies and $107$~kpc for blue galaxies,
but the behavior of $R_{\ast}$ with color is similar in that our red
galaxies have larger $R_{\ast}$ than blue galaxies.

The $R(L)$ dependence on galaxy color is similar in the $K$-band
(bottom panels of Figure~\ref{fig:typeenv}). The halo absorption
radius, $R_{\ast}$, is again larger for red galaxies (though the
values are still consistent within uncertainties), with $R_{\ast} \sim
75$~kpc and $R_{\ast} \sim 105$~kpc for blue and red galaxies,
respectively. The luminosity scaling is steeper in red galaxies,
$\beta \sim 0.5$, than blue galaxies, $\beta \sim 0.2$, and the
covering fraction is larger in blue galaxies than red galaxies
(${\frl} \sim 0.9$ and ${\frl} \sim 0.6$, respectively).

\begin{deluxetable}{ccccccc}
\tablecolumns{7}
\tablewidth{0pt}
\tablecaption{Luminosity Scaling for Subsamples at $W_{\rm cut}=0.3$~{\AA}\label{tab:BKzhalos}}
\tablehead{
\colhead{\phantom{x}}       &
\multicolumn{3}{c}{------------- $B$-band -------------} &
\multicolumn{3}{c}{------------- $K$-band -------------} \\
\colhead{Sample} & 
\colhead{$R_{\ast}$}          & 
\colhead{$\beta$}            &  
\colhead{{\frl}}        &  
\colhead{$R_{\ast}$}          & 
\colhead{$\beta$}            &  
\colhead{{\frl}}        \\
\colhead{} & 
\colhead{[kpc]}          & 
\colhead{}            &  
\colhead{}        &  
\colhead{[kpc]}          & 
\colhead{}            &  
\colhead{}        
}
\startdata
All & $80^{+50}_{-8}$ & $0.30^{+0.03}_{-0.01}$ & $0.85^{+0.03}_{-0.04}$ & $75^{+40}_{-6}$ & $0.23^{+0.01}_{-0.01}$ & $0.84^{+0.04}_{-0.04}$ \\[3pt]
Blue & $87^{+28}_{-2}$ & $0.45^{+0.02}_{-0.13}$ & $0.86^{+0.05}_{-0.06}$ & $74^{+24}_{-1}$ & $0.17^{+0.04}_{-0.01}$ & $0.89^{+0.04}_{-0.06}$ \\[3pt]
Red & $110^{+33}_{-0}$ & $0.66^{+0.02}_{-0.13}$ & $0.64^{+0.02}_{-0.12}$ & $104^{+19}_{-13}$ & $0.46^{+0.14}_{-0.02}$ & $0.62^{+0.07}_{-0.07}$ \\[3pt]
Low z & $80^{+38}_{-2}$ & $0.38^{+0.08}_{-0.04}$ & $0.75^{+0.06}_{-0.07}$ & $78^{+32}_{-1}$ & $0.27^{+0.01}_{-0.04}$ & $0.78^{+0.06}_{-0.07}$ \\[3pt]
High z & $86^{+30}_{-1}$ & $0.27^{+0.07}_{-0.02}$ & $0.85^{+0.05}_{-0.06}$ & $82^{+35}_{-4}$ & $0.26^{+0.01}_{-0.06}$ & $0.83^{+0.05}_{-0.07}$ \\[-5pt]
\enddata
\end{deluxetable}

\subsubsection{Galaxy Redshift}

We also examined the $R(L)$ dependence in both the $B$- and $K$-bands
on galaxy redshift, slicing the sample into low and high redshift
galaxies by the median redshift $\langle z\rangle =0.359$, and
adopting $W_{\rm cut}=0.3$~{\AA}. Values for $R_{\ast}$, $\beta$, and
${\frl}$ are presented in Figure~\ref{fig:DLz} and tabulated in
Table~\ref{tab:BKzhalos}.

In the $B$-band (top panels of Figure~\ref{fig:DLz}), we find
$R_{\ast} \sim 80$~kpc, $\beta \sim 0.4$, and ${\frl} \sim 0.75$ in
the low $z$ subsample. The high $z$ subsample has similar values for
$R_{\ast}$ ($\sim 85$~kpc) and ${\frl}$ ($\sim 0.8$), but a much
shallower dependence on $L_B/L_B^{\ast}$, with $\beta \sim 0.3$. This
difference in $\beta$ may be due to selection effects in that the
lowest luminosity galaxies ($L_B/L_B^{\ast}<0.1$) are being selected
against at high redshift. The selection methods for {\magiicat}
galaxies are detailed in Paper I \citep{nielsen12a}. Additionally, the
presence of galaxies at low z and low impact parameter may bias
$\beta$ to a larger value.

On the other hand, the $K$-band (bottom panels of
Figure~\ref{fig:DLz}) shows no discernable difference within
uncertainties between the low and high $z$ subsamples. The values for
$R_{\ast}$, $\beta$, and ${\frl}$ for both subsamples are $\sim
80$~kpc, $\sim 0.3$, and $\sim 0.8$, respectively. Comparing to the
apparent $B$-band evolution, this may imply that the luminosity
dependence of the extent of the {\MgII} absorbing CGM is unchanged
with redshift for the $K$-band. An alternate explanation is that the
$K$-band is not impacted by selection effects as much as the $B$-band
since $K$-band absolute magnitudes and luminosities were calculated
from apparent $K$-band magnitudes for half of the sample. The $K$-band
apparent magnitude is not one of the galaxy selection criteria
\citep{nielsen12a}.

\section{Discussion}
\label{sec:discussion}

With the larger sample afforded by {\magiicat}, we have obtained
additional leverage to probe the dependence of the properties of the
{\MgII} absorbing CGM with galaxy color, redshift, luminosity, impact
parameter, and $W_r(2796)$ threshold. For our analysis, we have
purposely avoided couching the data within the framework of model
expectations or scaling various measured quantities with ``second
parameters'', such as scaling the impact parameter by galaxy
luminosity, in order to provide a direct view of the CGM-galaxy
connection and to interpret the data from a purely observational
perspective.

We confirm the well-known anti-correlation between {\MgII} equivalent
width, $W_r(2796)$, and impact parameter, $D$, which is significant to
the $7.9~\sigma$ level. We find that the general behavior of this
anti-correlation on the $W_r(2796)-D$ plane is best fit by a
log-linear fit to the data, $\log W_r(2796)=\alpha_1D+\alpha_2$, where
$\alpha_1=-0.015 \pm 0.002$ and $\alpha_2=0.27 \pm 0.11$. This is in
contrast to the power law fit obtained by \citet{chen10a}. Our
log-linear fit predicts a leveling off of $W_r(2796)$ as $D$ goes to
zero, whereas the \citet{chen10a} fit does not. In fact, for
$D=5$~kpc, which is roughly the smallest impact parameter in
{\magiicat}, our log-linear fit predicts $W_r(2796)\simeq 1.6$~{\AA},
while the power law predicts $W_r(2796)\simeq 4.4$~{\AA} (and
continues to increase, for example at $D=1$~kpc, the power law fit
predicts $W_r(2796)\simeq 30$~{\AA}). It would seem reasonable that
the equivalent width distribution should flatten as one probes galaxy
disks as $W_r(2796) > 10$~{\AA} has yet to be reported in the
literature, even for the Milky Way interstellar medium.

As many previous works have noted (see \S~\ref{sec:Wr-D} for
references), there is considerable scatter in the $W_r(2796)-D$ plane,
which we also observe in Figure~\ref{fig:EWvsD}. Slicing {\magiicat}
by the median values of $B-K$, $z_{\rm gal}$, $L_B/L_B^{\ast}$, and
$L_K/L_K^{\ast}$ (see Figure~\ref{fig:EWD2cuts}), we find that the
scatter is not related to galaxy color or redshift. However, we do
find that the scatter may be related to galaxy luminosities such that
higher luminosity galaxies systematically populate the $W_r(2796)-D$
plane at larger $W_r(2796)$ and larger $D$. Performing a 2DKS test on
luminosities in the $W_r(2796)-D$ plane, we find this result to be
significant to the $4.2~\sigma$ level for the $B$-band and to the
$4.3~\sigma$ level for the $K$-band. Given that the $K$-band can be
considered a proxy for stellar mass, this result may indicate that the
scatter in the $W_r(2796)-D$ plane is actually due to galaxy mass such
that more massive galaxies have larger $W_r(2796)$ for a given $D$. In
fact, \citet{churchill-masses} find this to be true for {\magiicat}
galaxies with the galaxy virial mass obtained using halo abundance
matching.

Alternatively, the luminosity segregation on the $W_r(2796)-D$ plane
may be due to a gas metallicity-luminosity correlation. For damped
{\Lya} systems, \citet{turnshek05} found that gas metallicity
correlates with {\MgII} equivalent width. If higher luminosity
galaxies have more metal-rich gas, then $W_r(2796)$ would tend to be
larger than for lower luminosity galaxies, and the gas would be
detectable to larger impact parameters. This behavior on the
$W_r(2796)-D$ plane might have some relationship to the observed
metallicity bimodality in Lyman limit systems \citep{lehner13}.

With regard to the relationship between galaxy color and $W_r(2796)$,
our results challenge those of previous studies. For the overall
sample, we find no correlation between $W_r(2796)$ and $B-K$.
However, we find that for the subsample with $W_r(2796) \geq 1$~{\AA},
redder galaxies have larger $W_r(2796)$ with a significance level of
$2.5~\sigma$. These findings are contrary to the results of
\citet{zibetti07} and \citet{bordoloi11}, who both report larger
$W_r(2796)$ associated with bluer galaxies. It is difficult to rectify
the inconsistencies of our findings with those works unless the
solution originates in the different experimental methods.

\citet{zibetti07} developed an image stacking method in which the
individual galaxies are not identified, but light from galaxies in
various SDSS photometric bands are measured {\it relative to control
  fields\/} for which the quasars show no {\MgII} absorption.  The
color cuts of the galaxies studied by \citet{bordoloi11} are based
upon a color-mass relation using $u-B$ colors. Furthermore, the
equivalent widths measured by \citet{bordoloi11} are based on stacked
spectra over fixed impact parameter annuli in which the {\MgII}
doublet is not resolved, whereas those measured in this work are based
upon pencil-beam probes of the CGM for which all doublets are cleanly
resolved. If the CGM is intrinsically patchy
\citep[see][]{churchill-weakgals} with covering fraction dependencies
on $W_r(2796)$ threshold and galaxy luminosity (as we have
demonstrated in this paper), then the two methods (statistical versus
case-by-case) may be providing different but complimentary clues to
the nature of the CGM.

Our analysis of the covering fraction profile of the {\MgII} absorbing
CGM as a function of impact parameter, galaxy luminosity, and
absorption strength threshold provides improved insights on the extent
and distribution of the low ionization, metal-enriched CGM (see
Figure~\ref{fig:fdbar} and Table~\ref{tab:fdbar}). The decrease in the
covering fraction with increasing absorption threshold indicates that
the cross-section of the highest column density and/or most
kinematically dispersed gas (plausibly winds material) is smaller than
the cumulative cross-section including lower column density,
kinematically quiescent gas (plausibly accretion material).

Additional insight is provided by the relative behavior of the
covering fraction profile of low and high luminosity galaxies
(bifurcated at $L_B/L_B^{\ast}=0.611$). First, the {\MgII} absorbing
CGM appears to extend no further than 100~kpc for low luminosity
galaxies. Second, the CGM at $D \geq 100$~kpc surrounding high
luminosity galaxies is dominated by lower column density,
kinematically quiescent gas traced by $W_r(2796) < 0.6$~{\AA}
absorption; the cross-section of this material increases as the
equivalent width threshold is decreased to 0.1~{\AA}.  In contrast,
within $D<100$~kpc, the observed frequencies of lower and higher
column density CGM gas is not strongly dependent on galaxy luminosity
for $W_{\rm cut} = 0.1$, 0.3, and 0.6~{\AA}, but for $W_{\rm cut} =
1.0$~{\AA}, high luminosity galaxies have a higher observed frequency
of {\MgII} absorbing CGM. For $D \leq 25$~kpc, the covering fraction
is effectively unity for the thresholds $W_r(2796) \geq 0.1$ and
0.3~{\AA}, but declines for $W_r(2796) \geq 0.6$ and 1.0~{\AA} as the
threshold is increased.

These results indicate a strong luminosity dependence in
cross-sections and extent of the {\MgII} absorbing CGM such that high
luminosity galaxies have both a much more extended diffuse CGM than
low luminosity galaxies and higher filling factors of the highest
column density material within 50~kpc as compared to low luminosity
galaxies. Since the luminosity segregation we are employing is for the
$B$-band, we could be seeing a connection between the CGM and young
stars, such that the increased presence of young stars is related to a
greater cross-section of higher column density gas in the inner
regions of the galaxies and a more extended distribution of the lower
column density gas in the outer regions of the galaxies (beyond
100~kpc). This would imply a stellar driven mechanism for distributing
the gas further out into the CGM. 

We also find a difference in the covering fraction as a function of
galaxy color outside of 100~kpc. In this case, blue galaxies have
higher covering fractions than red galaxies, which have no absorption
beyond 100~kpc for the highest column density material ($W_r(2796)\geq
0.6$~{\AA}). Blue galaxies are more likely to host the galactic-scale
winds which lift more metal-rich material into the halo. Therefore,
the larger covering fractions in blue galaxies could be due to more
metal-rich material present at large impact
parameters. \citet{lehner13} concluded that Lyman limit systems, which
trace the cool CGM, have a bimodal metallicity distribution where the
metal-poor branch is likely probing cold accretion streams while the
metal-rich branch could be tracing recycled outflowing
winds. Therefore, our implied stellar driven mechanism may be driving
more metal-rich material into the CGM. This metal-rich material may
result in larger $W_r(2796)$, which in turn could cause higher
covering fractions for more luminous galaxies at large impact
parameters.

Following previous studies, we have assumed the relation $R(L) =
R_{\ast} (L/L^{\ast})^{\beta}$ to describe the ``halo absorption
radius'' dependence on luminosity. For the $B$-band luminosity, we
find that the sensitivity of $R(L)$ to luminosity (the parameter
$\beta$, see Table~\ref{tab:halos} and Figure~\ref{fig:DLenv})
increases as the equivalent width threshold is raised. For $W_r(2796)
\geq 0.1$~{\AA}, we find $\beta \simeq 0.3$, and for $W_r(2797) \geq
0.3$, 0.6, and 1.0~{\AA}, we find $\beta \simeq 0.4$. The extent for
an $L_B^{\ast}$ galaxy decreases with increasing equivalent width
threshold. We find $R_{\ast} \simeq 80$~{kpc} for $W_r(2796) \geq 0.1$
and 0.3~{\AA}, and $R_{\ast} \simeq 75$~{kpc} for $W_r(2796) \geq 0.6$
and 1.0~{\AA}, though these values are consistent within
uncertainties. In the $K$-band, we find similar values of $R_{\ast}$
($\sim 80$~kpc) for $W_{\rm cut}=0.1$, 0.3, and 0.6~{\AA}, while the
most optically thick gas has a much smaller absorption radius
($R_{\ast} \sim 60$~kpc for $W_{\rm cut}=1.0$~{\AA}). Furthermore, we
find a shallower dependence of $R(L)$ on the $K$-band luminosity,
where $\beta$ ranges from $\sim 0.2$ to $\sim 0.3$.  The greater
extent in the most optically thick gas for the $B$-band compared to
the $K$-band and the higher luminosity sensitivity to the $B$-band
relative to the $K$-band would further strengthen the idea that the
geometric extent and morphology of the {\MgII} absorbing CGM is
governed in large part by young stars.

There is an indication that galaxy color directly plays a role in
governing the luminosity sensitivity of $R(L)$. For $W_r(2796) \geq
0.3$~{\AA}, red galaxies have a remarkably steeper luminosity
dependence than do blue galaxies in the $B$-band. For red galaxies, we
find $\beta = 0.66$, whereas for blue galaxies, we find $\beta =
0.45$. In the $B$-band, the covering fraction within $R(L)$ of red
galaxies is $\simeq 30$\% lower than for blue galaxies. For the
$K$-band, red galaxies still have a much steeper luminosity dependence
($\beta = 0.46$) than blue galaxies ($\beta = 0.17$), as well as a
$\sim 40\%$ smaller covering fraction. In studying the $K$-band of
Figure~\ref{fig:typeenv}, it is apparent that red galaxies are, on
average, brighter in the $K$-band than blue galaxies. Using the
$K$-band luminosity as a proxy for stellar mass, this indicates that
red galaxies are, on average, more massive than blue
galaxies. Therefore, we may be seeing that the halo absorption radius
depends on the mass of the galaxy such that more massive galaxies have
larger halo absorption radii resulting in a larger $\beta$ for redder,
more massive galaxies. The dependence of the halo absorption radius on
mass is explored in more depth in the third paper of this series
\citep{churchill-masses2}.

We strongly caution that the function $R(L)$ should not be viewed as a
well-defined outer boundary to the {\MgII} absorbing CGM. The values
of $R_{\ast}$ and $\beta$ are dependent on the sample and may change
if we obtain more low luminosity galaxies with $L/L^{\ast}<0.1$. We
are confident, however, that larger luminosity galaxies have a more
extended CGM, i.e., the value of $\beta$ is positive for both the $B$-
and $K$-bands. Additionally, examination of the data on the
$D$--$L/L^{\ast}$ plane shown in Figure~\ref{fig:DLenv} clearly
reveals that a non-negligible number of galaxies with $W_r(2796)$
greater than the threshold cut also reside above the curve. The
$1~\sigma$ upper envelope of the $R(L)$ boundary characterizes the
presence of these galaxies. These galaxies also drive the large upward
uncertainties in the best fit $R_{\ast}$. Note that this uncertainty
increases from $\simeq 26$ to $\simeq 50$~kpc as the absorption
threshold is decreased from $W_r(2796) \geq 1.0$ to 0.1~{\AA}. This
suggests that the $R(L)$ boundary is less well defined for the lower
column density structures of the CGM. Physically, the behavior of the
data in this regime of impact parameter and $W_r(2796)$ could either
reflect a greater level of patchiness in the low column density
material residing in the outskirts of the CGM of individual galaxies
or a broad range of CGM properties from galaxy to galaxy.

\section{Summary}
\label{sec:concl}

Combining our previous studies and the extant works in the literature,
we have compiled a ``{\MgII} Absorber-Galaxy Catalog'' ({\magiicat})
of intermediate redshift galaxies and their associated circumgalactic
medium (CGM) as probed using {\MgIIdblt} absorption. Details of the
{\magiicat} data are presented in Paper I \citep{nielsen12a}. In this
paper, we present results from a first analysis in which we compare
only direct observables and avoid converting or scaling the data to a
preferred CGM model, focusing exclusively on ``isolated'' galaxies,
which are defined to have no neighboring galaxy within a projected
distance of 100~kpc and line of sight velocity within 500~{\kms}.

The sample presented here comprises 182 galaxies toward 134 quasar
sightlines over the redshift range $0.072 \leq z_{\rm gal} \leq
1.120$, with median $\langle z \rangle = 0.359$. The rest-frame
magnitudes range from $-16.1 \geq M_B \geq -23.1$ ($0.02 \leq
L_B/L_B^{\ast} \leq 5.87$) and $-17.0 \geq M_K \geq -25.3$ ($0.006 \leq
L_K/L_K^{\ast} \leq 9.71$) with AB rest-frame colors $0.04 \leq B-K
\leq 4.09$. The median $B$-band luminosity is $L_B/L_B^{\ast}=0.611$
and the median color is $B-K=1.48$.

The main results are:

1. The mean $M_B$ increases with increasing redshift ($4.4~\sigma$),
whereas $M_K$ exhibits a weak trend to increase with redshift
($2.2~\sigma$). Galaxy luminosities $L_B/L_B^{\ast}$ and
$L_K/L_K^{\ast}$ do not evolve with redshift. The rest-frame $B-K$
color shows no redshift evolution, consistent with the findings of
\citet{zibetti07}.

2. There is no correlation between $W_r(2796)$ and $B-K$ for the full
sample ($1.3~\sigma$). However, for galaxies associated with
$W_r(2796) \geq 1.0$~{\AA} absorption, $B-K$ correlates with
$W_r(2796)$ at $2.5~\sigma$, indicating a trend that redder galaxies
have stronger absorption in this equivalent width regime. In our
sample, the distributions of $W_r(2796)$ within $D=50$~kpc of blue
($B-K < 1.48$) and red ($B-K \geq 1.48$) galaxies are consistent with
being drawn from the same parent distribution ($0.3~\sigma$). The
$W_r(2796)$-color correlation in our sample conflicts with the
$W_r(2796)$-color anti-correlation reported by \citet{zibetti07}.
Both the $W_r(2796)$-color correlation and the indistinguishable
$W_r(2796)$ distributions in our sample are in conflict with the
findings of \citet{bordoloi11}, who report eight times stronger
$W_r(2796)$ within $D=50$~kpc associated with bluer galaxies.

3. Including upper limits on $W_r(2796)$, we find that the
significance level of the well-known anti-correlation between
$W_r(2796)$ and impact parameter, $D$, is $7.9~\sigma$ for this
sample. The best parameterization is a log-linear relation, for which
we find $\log W_r(2796) = (-0.015 \pm 0.002)D + (0.27 \pm 0.11)$. The
scatter in this relation may be due to the redshifts, colors, or
luminosities of the galaxies. Splitting the sample by the $K$-band
luminosity yields a $4.3~\sigma$ significance that low and high
$L_K/L_K^{\ast}$ galaxies are separate populations on the
$W_r(2796)-D$ plane such that higher $K$-band luminosity galaxies tend
to have higher $W_r(2796)$ at a given impact parameter. Dividing the
sample into quartiles based on the redshift, color, or luminosity of
the galaxies and comparing the lowest and highest quartiles yields at
best a $3.4~{\sigma}$ significance with $K$-band luminosity.

4. The covering fraction profiles with projected distance from the
galaxy, $\fdmax$ and $\fdbar$, decrease with both increasing $D$ and
increasing minimum $W_r(2796)$ (see Figures~\ref{fig:fdmax} and
\ref{fig:fdbar}). High luminosity galaxies have higher covering
fractions. In terms of $\fdbar$, this difference is greatest at $100 <
D \leq 200$~kpc for absorbing CGM gas with $W_r(2796) \geq 0.1$~{\AA}
and at $D \leq 50$~kpc for CGM gas with $W_r(2796) \geq
1.0$~{\AA}. There is no clear difference between the covering fraction
profile of blue galaxies versus red galaxies within 100~kpc, though
the gas in red galaxies may be more concentrated near the center of
the galaxy than in blue galaxies. Outside 100~kpc, red galaxies have
smaller covering fractions than blue galaxies and no absorption for
$W_r(2796) \geq 0.6$. The high redshift galaxies ($z \geq \langle z
\rangle$) have a higher covering fraction than the low redshift
galaxies ($z < \langle z \rangle$) at $D>50$~kpc.

5. We determined the best-fit parameters $R_{\ast}$ and $\beta$ for
the luminosity scaling of the outer envelope for absorption, or the
``halo absorption radius'', $R(L) = R_{\ast}(L/L^{\ast})^{\beta}$, for
both the $B$- and $K$-bands as a function of $W_r(2796)$ threshold.
For the $B$-band, the luminosity scaling increases from $\beta \sim
0.3$ to $\beta \sim 0.4$ with increasing $W_{\rm cut}$. We find
$R_{\ast} \sim 80$~kpc for $W_{\rm cut}=0.1$ and 0.3~{\AA}, and
$R_{\ast} \sim 75$~kpc for $W_{\rm cut}=0.6$ and 1.0~{\AA}. The
covering fraction inside $R(L_B)$ decreases from ${\frl}=0.85$ for
$W_{\rm cut}=0.1$~{\AA} to ${\frl}=0.34$ for $W_{\rm
  cut}=1.0$~{\AA}. For the $K$-band, the luminosity scaling ranges
from $\beta \sim 0.2$ to 0.3, and $R_{\ast}$ ranges from $\sim 60$~kpc
to $\sim 80$~kpc. The covering fraction behaves similarly to the
$B$-band. Though the ``outer envelope'' for absorption has a clear
dependence on both luminosity and $W_r(2796)$ threshold, we note that
the scatter in the data in the $W_r(2796)$-$D$ plane, the behavior of
the covering fraction profile $\fdbar$ with increasing $D$, and the
relatively extended upward uncertainties in $R^{\ast}$ all indicate
that $R(L)$ is not a well-defined quantity, but should be interpreted
as a ``fuzzy'' boundary and that the ``fuzziness'' increases with
decreasing equivalent width threshold.

6. Dividing the galaxies by the median color ($B-K = 1.48$), adopting
$W_{\rm cut}=0.3$~{\AA}, and examining the fit parameters to
$R(L)=R_{\ast}(L/L^{\ast})^{\beta}$, we find that red galaxies have a
steeper luminosity scaling, $\beta$, and larger halo absorption radii,
$R_{\ast}$, than blue galaxies in both the $B$- and $K$-bands.  This
behavior of larger $R_{\ast}$ for red galaxies is consistent with the
results of \citet{bordoloi11}, though our values are smaller. Using
the $K$-band as a proxy for stellar mass, we may be seeing that the
difference in the values of $R_{\ast}$ and $\beta$ between red and
blue galaxies is due to the mass of a galaxy such that more massive
galaxies have larger halo absorption radii \citep[also
  see][]{churchill-masses2}.

7. We also examined the $R(L)$ dependence in both the $B$- and
$K$-bands on galaxy redshift, slicing the sample at the median
redshift $z=0.359$, and adopting $W_{\rm cut}=0.3$~{\AA}. In the
$B$-band, we find $R_{\ast} \sim 80$~kpc, $\beta \sim 0.4$, and
${\frl} \sim 0.8$ in the low $z$ subsample. The high $z$ subsample has
similar values for $R_{\ast}$ ($\sim 85$~kpc) and ${\frl}$ ($\sim
0.8$), but a much shallower dependence on $L_B/L_B^{\ast}$, with
$\beta \sim 0.3$. In the $K$-band, we find no discernable difference
within uncertainties between the low and high $z$ subsamples. The
values for $R_{\ast}$, $\beta$, and ${\frl}$ for both subsamples are
$\sim 80$~kpc, $\sim 0.3$, and $\sim 0.8$, respectively.

This work constitutes our first examination of the relationship of the
{\MgII} absorbing CGM with galaxy properties using {\magiicat}. In
particular, we have examined the data by comparing direct observables
and have avoided analyzing the quantities based upon converting or
scaling the data to a preferred model of the CGM. Our aim has been to
provide an unfiltered view of the {\MgII} absorbing CGM. Nonetheless,
the simple bivariate testing we have conducted here does not provide
full leverage for probing of the available data. In future work, we
will apply multivariate analysis methods in which we also incorporate
mass estimates of the galaxies, {\MgII} kinematics, and both low and
high ionization absorption strengths of the CGM for these galaxies. We
will also conduct comparative studies of the isolated galaxies and
group galaxies in the catalog.

\acknowledgments 

We thank the anonymous referee for a careful reading of the manuscript
and for valuable input that improved this work.  We also thank Nick
Gnedin for reading and providing insightful comments on an earlier
draft. This research was primarily supported through grant
HST-AR-12646 provided by NASA via the Space Telescope Science
Institute, which is operated by the Association of Universities for
Research in Astronomy (AURA) under NASA contract NAS 5-26555. This
work was also supported by the Research Enhancement Program provided
by NASA's New Mexico Space Grant Consortium (NMSGC). NMN was also
partially supported through a NMSGC Graduate Fellowship and a
three-year Graduate Research Enhancement Grant (GREG) sponsored by the
Office of the Vice President for Research at New Mexico State
University.


\begin{thebibliography}{}

\bibitem[Adelberger {\etal}(2005)]{adelberger05} Adelberger, K.~L.,
  Shapley, A.~E., Steidel, C.~C., et al.\ 2005, \apj, 629, 636

\bibitem[Barton \& Cooke(2009)]{barton09} Barton, E.~J., \& Cooke,
  J.\ 2009, \aj, 138, 1817

\bibitem[Bechtold \& Ellingson(1992)]{bechtold92} Bechtold, J., \&
  Ellingson, E.\ 1992, \apj, 396, 20

\bibitem[Benjamin \& Danly(1997)]{benjamin97} Benjamin, R.~A., \&
  Danly, L.\ 1997, \apj, 481, 764

\bibitem[Bergeron \& Boiss\`{e}(1991)]{bb91} Bergeron, J., \&
  Boiss\`{e}, P. 1991, A\&A, 243, 334

\bibitem[Bergeron \& Stasi{\'n}ska(1986)]{bs86} Bergeron, J., \&
  Stasi{\'n}ska, G.\ 1986, \aap, 169, 1

\bibitem[Birnboim \& Dekel(2003)]{birnboim03} Birnboim, Y., \& Dekel,
  A.\ 2003, MNRAS, 345, 349

\bibitem[Birnboim {\etal}(2007)]{birnboim07} Birnboim, Y., Dekel, A.,
  \& Neistein, E.\ 2007, MNRAS, 380, 339

\bibitem[Bolzonella {\etal}(2000)]{bolzonella00} Bolzonella, M.,
  Miralles, J.-M., \& Pell{\'o}, R.\ 2000, \aap, 363, 476

\bibitem[Bordoloi {\etal}(2011)]{bordoloi11} Bordoloi, R., Lilly, S.~J.,
  Knobel, C., {\etal} 2011, \apj, 743, 10

\bibitem[Bouch{\'e} {\etal}(2012)]{bouche11} Bouch{\'e}, N., Hohensee,
  W., Vargas, R., {\etal}\ 2012, \mnras, 426, 801

\bibitem[Bouch{\'e} {\etal}(2006)]{bouche06} Bouch{\'e}, N., Murphy,
  M.~T., P{\'e}roux, C., Csabai, I., \& Wild, V. 2006, \mnras, 371,
  495

\bibitem[Brown, Hollander, \& Korwar(1974)]{bhk} Brown, B.~W.,
  Hollander, M., \& Korwar, R.~M. 1974, in Reliability and Biometry,
  327

\bibitem[Bruzual \& Charlot(1993)]{bc93} Bruzual A.~G., \&
  Charlot, S.\ 1993, \apj, 405, 538

\bibitem[Charlton \& Churchill(1996)]{cc96} Charlton, J.~C., \&
  Churchill, C.~W.\ 1996, \apj, 465, 631

\bibitem[Chelouche \& Bowen(2010)]{chelouche10} Chelouche, D., \&
  Bowen, D.~V.\ 2010, \apj, 722, 1821

\bibitem[Ceverino \& Klypin(2009)]{ceverino09} Ceverino, D., \&
  Klypin, A.\ 2009, \apj, 695, 292

\bibitem[Charlton \& Churchill(1998)]{cc98} Charlton, J.~C., \&
  Churchill, C.~W.\ 1998, \apj, 499, 181

\bibitem[Chen {\etal}(2010a)]{chen10a} Chen, H.-W., Helsby, J.~E.,
  Gauthier, J.-R., Shectman, S.~A., Thompson, I.~B., \& Tinker,
  J.~L.\ 2010a, ApJ, 714, 1521

\bibitem[Chen {\etal}(2001a)]{chen01a} Chen, H.-W.,
  Lanzetta, K.~M., \& Webb, J.~K.\ 2001a, \apj, 556, 158

\bibitem[Chen {\etal}(2001b)]{chen01b} Chen, H.-W., Lanzetta, K.~M.,
  Webb, J.~K., \& Barcons, X. 2001b, ApJ, 559, 654

\bibitem[Chen {\etal}(2010b)]{chen10b} Chen, H.-W., Wild, V., Tinker,
  J.~L., {\etal} 2010b, \apjl, 724, L176

\bibitem[Chen \& Tinker(2008)]{chen08} Chen, H.-W., \& Tinker,
  J.~L.\ 2008, ApJ, 687, 745

\bibitem[Churchill {\etal}(2013a)]{churchill-weakgals} Churchill,
  C.~W., Kacprzak, G.~G., Nielsen, N.~M., Steidel, C.~C., Murphy,
  M.~T. 2013a, ApJ, submitted

\bibitem[Churchill {\etal}(2005)]{cwc-china} Churchill, C.~W.,
  Kacprzak, G.~G., \& Steidel, C.~C.\ 2005, IAU Colloq.~199: Probing
  Galaxies through Quasar Absorption Lines, 24

\bibitem[Churchill {\etal}(2007)]{cwc1317} Churchill, C.~W., Kacprzak,
  G.~G., Steidel, C.~C., \& Evans, J.~L.\ 2007, \apj, 661, 714

\bibitem[Churchill {\etal}(2000a)]{archiveI} Churchill, C.~W., Mellon,
  R.~R., Charlton, J.~C., Jannuzi, B.~T., Kirhakos, S., Steidel,
  C.~C., \& Schneider, D.~P. 2000a, ApJS, 130, 91

\bibitem[Churchill {\etal}(2000b)]{archiveII} Churchill, C.~W., Mellon,
  R.~R., Charlton, J.~C., {\etal} 2000b, \apj, 543, 577

\bibitem[Churchill {\etal}(2013b)]{churchill-masses} Churchill, C.~W.,
  Nielsen, N.~M., Kacprzak, G.~G., \& Trujillo-Gomez, S.\ 2013b,
  \apjl, 763, L42

\bibitem[Churchill {\etal}(1999)]{weakI} Churchill, C.~W., Rigby,
  J.~R., Charlton, J.~C., \& Vogt, S.~S.\ 1999, \apjs, 120, 51

\bibitem[Churchill {\etal}(1996)]{csv96} Churchill, C.~W., Steidel,
  C.~C., \& Vogt, S.~S.\ 1996, \apj, 471, 164

\bibitem[Churchill {\etal}(2013c)]{churchill-masses2} Churchill,
  C.~W., Trujillo-Gomez, S., Nielsen, N.~M., \& Kacprzak,
  G.~G.\ 2013c, \apj, submitted

\bibitem[Churchill \& Vogt(2001)]{cv01} Churchill, C.~W., \& Vogt,
  S.~S.\ 2001, \aj, 122, 679

\bibitem[Churchill {\etal}(2003)]{cvc03} Churchill, C.~W., Vogt, S.~S.,
  \& Charlton, J.~C.\ 2003, \aj, 125, 98

\bibitem[Cirasuolo {\etal}(2010)]{cirasuolo} Cirasuolo, M., McLure,
  R.~J., Dunlop, J.~S., Almaini, O., Foucaud, S., \& Simpson, C. 2010,
  MNRAS, 401, 1166

\bibitem[Coleman {\etal}(1980)]{cww80} Coleman, G.~D., Wu, C.-C.,
  \& Weedman, D.~W.\ 1980, \apjs, 43, 393

\bibitem[Dekel \& Birnboim(2006)]{dekel06} Dekel, A., \& Birnboim,
  Y.\ 2006, MNRAS, 368, 2

\bibitem[Dekel \& Silk(1986)]{dekel86} Dekel, A., \& Silk, J.\ 1986,
  \apj, 303, 39

\bibitem[Evans(2011)]{evans-thesis} Evans, J.~L. 2011, Ph.D.~Thesis,
  New Mexico State University

\bibitem[Faber {\etal}(2007)]{faber} Faber, S.~M., et al.\ 2007, ApJ,
  665, 265

\bibitem[Fried {\etal}(2001)]{fried01} Fried, J.~W., von Kuhlmann, B.,
  Meisenheimer, K., {\etal} 2001, \aap, 367, 788

\bibitem[Gauthier {\etal}(2009)]{gauthier09} Gauthier, J.-R., Chen,
  H.-W., \& Tinker, J.~L. 2009, \apj, 702, 50

\bibitem[Gehrels(1986)]{gehrels86} Gehrels, N. 1986, ApJ, 303, 336

\bibitem[Guillemin \& Bergeron(1997)]{guillemin} Guillemin, P., \&
  Bergeron, J. 1997, A\&A, 328, 499

\bibitem[Kacprzak \& Churchill(2011a)]{ggk-omegaHI} Kacprzak, G.~G., \&
  Churchill, C.~W.\ 2011a, \apjl, 743, L34

\bibitem[Kacprzak {\etal}(2011b)]{kacprzak11kin} Kacprzak, G.~G.,
  Churchill, C.~W., Barton, E.~J., \& Cooke, J.\ 2011b, ApJ, 733, 105

\bibitem[Kacprzak {\etal}(2010)]{ggk-sims} Kacprzak, G.~G.,
  Churchill, C.~W., Ceverino, D., et al.\ 2010, \apj, 711, 533

\bibitem[Kacprzak {\etal}(2011c)]{kacprzak} Kacprzak, G.~G.,
  Churchill, C.~W., Evans, J.~L., Murphy, M.~T., \& Steidel,
  C.~C.\ 2011c, MNRAS, 416, 3118

\bibitem[Kacprzak, Churchill, \& Nielsen(2012)]{kcn} Kacprzak, G.~G.,
  Churchill, C.~W., \& Nielsen, N.~M. 2012, ApJ, 760, L7

\bibitem[Kacprzak {\etal}(2008)]{ggk08} Kacprzak, G.~G., Churchill,
  C.~W., Steidel, C.~C., \& Murphy, M.~T.\ 2008, \aj, 135, 922

\bibitem[Kacprzak {\etal}(2007)]{ggk-morphology} Kacprzak, G.~G.,
  Churchill, C.~W., Steidel, C.~C., Murphy, M.~T., \& Evans,
  J.~L.\ 2007, \apj, 662, 909

\bibitem[Kacprzak {\etal}(2012)]{ggk-1317} Kacprzak, G.~G., Churchill,
  C.~W., Steidel, C.~C., Spitler, L.~R., \& Holtzman, J.~A. 2012,
  MNRAS, 427, 3029

\bibitem[Kacprzak, Murphy, \& Churchill(2010)]{ggk1127} Kacprzak,
  G.~G., Murphy, M.~T., \& Churchill, C.~W. 2010, MNRAS, 406, 445

\bibitem[Kere{\v s} {\etal}(2009)]{keres09} Kere{\v s}, D., Katz, N.,
  Fardal, M., Dav{\'e}, R., \& Weinberg, D.~H.\ 2009, MNRAS, 395, 160

\bibitem[Kere{\v s} {\etal}(2005)]{keres05} Kere{\v s}, D., Katz, N.,
  Weinberg, D.~H., \& Dav{\'e}, R.\ 2005, MNRAS, 363, 2

\bibitem[Kim, Goobar, \& Perlmutter(1996)]{kim} Kim, A., Goobar, A.,
  \& Perlmutter, S. 1996, PASP, 108, 190

\bibitem[Lanzetta \& Bowen(1990)]{lanzetta90} Lanzetta, K.~M., \&
  Bowen, D.\ 1990, \apj, 357, 321

\bibitem[Lanzetta \& Bowen(1992)]{lb92} Lanzetta, K.~M., \& Bowen,
  D.~V.\ 1992, \apj, 391, 48

\bibitem[Lanzetta {\etal}(1995)]{lanzetta95} Lanzetta, K.~M., Bowen,
  D.~V., Tytler, D., \& Webb, J.~K. 1995, ApJ, 442, 538

\bibitem[Le Brun {\etal}(1993)]{lebrun93} Le Brun, V., Bergeron, J.,
  Boisse, P., \& Christian, C.\ 1993, \aap, 279, 33

\bibitem[Lehner {\etal}(2013)]{lehner13} Lehner, N., Howk, J.~C.,
  Tripp, T.~M., {\etal} 2013, \apj, 770, 138

\bibitem[Lilly {\etal}(1995)]{lilly95} Lilly, S.~J., Tresse, L.,
  Hammer, F., Crampton, D., \& Le Fevre, O.\ 1995, \apj, 455, 108

\bibitem[Lin {\etal}(1999)]{lin99} Lin, H., Yee, H.~K.~C., Carlberg,
  R.~G., {\etal}\ 1999, \apj, 518, 533

\bibitem[Lundgren {\etal}(2009)]{lundgren09} Lundgren, B.~F., Brunner,
  R.~J., York, D.~G., {\etal} 2009, \apj, 698, 819

\bibitem[Maller \& Bullock(2004)]{maller04} Maller, A.~H., \& Bullock,
  J.~S.\ 2004, \mnras, 355, 694

\bibitem[Martin \& Bouch{\'e}(2009)]{martin09} Martin, C.~L., \&
  Bouch{\'e}, N.\ 2009, \apj, 703, 1394

\bibitem[McGaugh {\etal}(2010)]{mcgaugh10} McGuagh, S.~S., Schombert,
  J.~M., de Blok, W.~J.~G., \& Zagursky, M.~J. 2010, ApJ, 708, L14

\bibitem[M{\'e}nard {\etal}(2011)]{menard11} M{\'e}nard, B., Wild, V.,
  Nestor, D., {\etal} 2011, \mnras, 417, 801

\bibitem[Mo \& Miralda-Escude(1996)]{mo96} Mo, H.~J., \&
  Miralda-Escude, J.\ 1996, \apj, 469, 589

\bibitem[Narayanan {\etal}(2007)]{anand07} Narayanan, A., Misawa, T.,
  Charlton, J.~C., \& Kim, T.-S.\ 2007, \apj, 660, 1093

\bibitem[Navarro {\etal}(1996)]{nfw96} Navarro, J.~F., Frenk, C.~S.,
  \& White, S.~D.~M.\ 1996, \apj, 462, 563

\bibitem[Nestor {\etal}(2005)]{nestor05} Nestor, D.~B., Turnshek,
  D.~A., \& Rao, S.~M.\ 2005, \apj, 628, 637

\bibitem[Nielsen {\etal}(2013)]{nielsen12a} Nielsen, N.~M., Churchill,
  C.~W., Kacprzak, G.~G., \& Murphy, M.~T.\ 2013, arXiv:1304.6716
  (Paper I)

\bibitem[Ocvirk, Pichon, \& Teyssier(2008)]{ocvirk08} Ocvirk, P.,
  Pichon, C., Teyssier, R. 2008, MNRAS, 390,1326

\bibitem[Oppenheimer {\etal}(2010)]{oppenheimer10} Oppenheimer, B.~D.,
  Dav{\'e}, R., Kere{\v s}, D., {\etal} 2010, MNRAS, 406, 2325

\bibitem[Petitjean \& Bergeron(1990)]{pb90} Petitjean, P., \&
  Bergeron, J.\ 1990, \aap, 231, 309

\bibitem[Rao \& Turnshek(2000)]{rao00} Rao, S.~M., \& Turnshek,
  D.~A.\ 2000, \apjs, 130, 1

\bibitem[Ribaudo {\etal}(2011)]{ribaudo11} Ribaudo, J., Lehner, N.,
  Howk, J.~C., {\etal} 2011, \apj, 743, 207

\bibitem[Rigby, Charlton, \& Churchill(2002)]{weakII} Rigby, J. R.,
  Charlton, J. C., \& Churchill, C. W. 2002, ApJ, 565, 743

\bibitem[Rubin {\etal}(2012)]{rubin11} Rubin, K.~H.~R., Prochaska,
  J.~X., Koo, D.~C., \& Phillips, A.~C.\ 2012, \apjl, 747, L26

\bibitem[Rubin {\etal}(2010)]{rubin10} Rubin, K.~H.~R., Weiner, B.~J.,
  Koo, D.~C., {\etal} 2010, \apj, 719, 1503

\bibitem[Rudie {\etal}(2012)]{rudie12} Rudie, G.~C., Steidel, C.~C.,
  Trainor, R.~F., {\etal} 2012, \apj, 750, 67

\bibitem[Schneider {\etal}(1993)]{schneider93} Schneider, D.~P.,
  {\etal} 1993, ApJS, 87, 45

\bibitem[Simcoe {\etal}(2006)]{simcoe06} Simcoe, R.~A., Sargent,
  W.~L.~W., Rauch, M., \& Becker, G.\ 2006, \apj, 637, 648

\bibitem[Steidel(1995)]{steidel95} Steidel, C.~C.\ 1995, QSO
  Absorption Lines, 139

\bibitem[Steidel {\etal}(1997)]{steidel97} Steidel, C.~C., Dickinson,
  M., Meyer, D.~M., Adelberger, K.~L., \& Sembach, K.~R. 1997, ApJ,
  480, 586

\bibitem[Steidel, Dickinson, \& Persson(1994)]{sdp94} Steidel, C.~C.,
  Dickinson, M., \& Persson, S.~E. 1994, ApJ, 437, L75

\bibitem[Steidel {\etal}(2010)]{steidel10} Steidel, C.~C., Erb, D.~K.,
  Shapley, A.~E., et al.\ 2010, \apj, 717, 289

\bibitem[Steidel \& Sargent(1992)]{ss92} Steidel, C.~C., \& Sargent,
  W.~L.~W. 1992, ApJS, 80, 1

\bibitem[Stewart {\etal}(2011)]{stewart11} Stewart, K.~R., Kaufmann,
  T., Bullock, J.~S., {\etal} 2011, ApJ, 738, 39

\bibitem[Tinker \& Chen(2008)]{tinkerchen08} Tinker, J.~L., \& Chen,
  H.-W.\ 2008, \apj, 679, 1218

\bibitem[Thom {\etal}(2011)]{thom11} Thom, C., Werk, J.~K., Tumlinson,
  J., {\etal} 2011, \apj, 736, 1

\bibitem[Tremonti {\etal}(2007)]{tremonti07} Tremonti, C.~A.,
  Moustakas, J., \& Diamond-Stanic, A.~M.\ 2007, \apjl, 663, L77

\bibitem[Tripp \& Bowen(2005)]{tripp-china} Tripp, T.~M., \& Bowen,
  D.~V. 2005, IAU Colloq.~199: Probing Galaxies through Quasar
  Absorption Lines, 5

\bibitem[Tumlinson {\etal}(2011)]{tumlinson11} Tumlinson, J., Thom,
  C., Werk, J.~K., {\etal} 2011, Science, 334, 948

\bibitem[Turnshek {\etal}(2005)]{turnshek05} Turnshek, D.~A., Rao,
  S.~M., Nestor, D.~B., Belfort-Mihalyi, M., \& Quider, A.~M.\ 2005,
  IAU Colloq.~199: Probing Galaxies through Quasar Absorption Lines,
  104

\bibitem[van de Voort {\etal}(2011)]{vandevoort11} van de Voort, F.,
  Schaye, J., Booth, C.~M., Haas, M.~R., \& Dalla Vecchia, C.\ 2011,
  MNRAS, 414, 2458

\bibitem[van de Voort \& Schaye(2012)]{vandevoort+schaye11} van de
  Voort, F., \& Schaye, J.\ 2012, \mnras, 423, 2991

\bibitem[Wang \& Wells(2000)]{wang} Wang, W., \& Wells, M.~T. 2000,
  Statistica Sinicia, 10, 1199

\bibitem[Weiner {\etal}(2009)]{weiner09} Weiner, B.~J., Coil, A.~L.,
  Prochaska, J.~X., et al.\ 2009, \apj, 692, 187

\bibitem[Weisheit(1978)]{weisheit78} Weisheit, J.~C.\ 1978, \apj, 219,
  829

\bibitem[Wolf {\etal}(2003)]{wolf03} Wolf, C., Meisenheimer, K., Rix,
  H.-W., {\etal} 2003, \aap, 401, 73

\bibitem[Wolynetz(1979)]{wolynetz79} Wolynetz, M. S. 1979, Journal of
  the Royal Statistical Society, 28, 195

\bibitem[Yanny \& York(1992)]{yanny92} Yanny, B., \& York,
  D.~G.\ 1992, \apj, 391, 569

\bibitem[Zibetti {\etal}(2007)]{zibetti07} Zibetti, S., M\`{e}nard, B.,
  Nestor, D.~B., Quider, A.~M., Rao, S.~M., \& Turnshek, D.~A. 2007,
  ApJ, 658, 161

\end{thebibliography}
\end{document}